\documentclass[useAMS,usenatbib]{mn2e}
\usepackage{graphicx,verbatim,amsfonts,amsmath,natbib}
\usepackage{bm}
\usepackage{times}

\newcommand{\hmpc}{h^{-1}{\rm\,Mpc}}

\newcommand{\ihmpcC}{h^3 {\rm\,Mpc}^{-3}}
\newcommand{\kmsmpc}{{\rm\ km\ s^{-1}\ Mpc^{-1}}}

\newcommand{\om}{\Omega_m}

\newcommand{\ol}{\Omega_\Lambda}

\newcommand{\sqdeg}{{\rm deg}^2}

\newcommand{\photo}{{\tt photo\,}}

\newcommand{\rpet}{\ensuremath{r_{\rm petro}}}

\newcommand{\clrg}{\ensuremath{c_{\rm LRG}}}

\newcommand{\etal}{et al.}
\newcommand{\beq}{\begin{equation}}
\newcommand{\eeq}{\end{equation}}
\newcommand{\beqa}{\begin{eqnarray}}
\newcommand{\eeqa}{\end{eqnarray}}

\newcommand{\ud}{\mathrm{d}}

\newcommand{\apj}{ApJ}
\newcommand{\aj}{AJ}

\newcommand{\apjl}{ApJL}
\newcommand{\mnras}{MNRAS}

\title{The Bright End of the Luminosity Function of Red Sequence Galaxies}
\author[Y.-S. Loh and M. A. Strauss]{Yeong-Shang Loh$^{1,2,3}$\thanks{E-mail:
yeongloh@colorado.edu} and Michael A. Strauss$^{1}$\\
$^{1}$Princeton University Observatory, Princeton, NJ 08544\\
$^{2}$Department of Physics, Princeton University, Princeton, NJ 08544\\
$^{3}$Center for Astrophysics and Space Astronomy, University of
Colorado, Boulder, CO 80309}
\begin{document}

\pagerange{\pageref{firstpage}--\pageref{lastpage}} \pubyear{0000}

\maketitle

\label{firstpage}

\begin{abstract}
We study the bright end of the luminosity distribution of galaxies in
fields with Luminous Red Galaxies (LRG) from the Sloan Digital Sky
Survey (SDSS). Using $2099\,\sqdeg$ of SDSS imaging data, we search for
luminous ($\ga L^*$) early-type galaxies within $1.0\,\hmpc$ of 
a volume-limited sample of $12,608$
spectroscopic LRG in the redshift range $0.12 < z < 0.38$. Most of
these objects lie in rich environments, with the LRG being the
brightest object within $1.0\,\hmpc$.  The luminosity gap, $M_{12}$,
between the first and second-rank galaxies within $1.0\,\hmpc$ 
is large ($\sim 0.8$ {\tt
  mag}), substantially larger than can be explained with an
exponentially decaying luminosity function of 
galaxies.  The brightest member is less luminous
(by 0.1 -- 0.2 {\tt mag}), and shows a larger gap in LRG selected groups than in
cluster-like environments. The large luminosity gap shows little 
evolution with redshift to $z = 0.4$,
ruling out the scenario that these LRG selected brightest cluster or group galaxies grow
by recent cannibalism of cluster members.
\end{abstract}

\begin{keywords}
methods: statistical -- galaxies: elliptical and lenticular, cD -- galaxies: evolution --
galaxies: clusters: general
\end{keywords}

\section{Introduction}\label{sec:gap_intro}
The bright end of the galaxy luminosity function is still not
completely characterized. While uncertainties about surface brightness 
completeness is an issue at the faint end \citep{Dal98},
the limiting factor at the bright end is often small number statistics. 
Early studies of
the bright end of the galaxy luminosity function focused on high
density regions, often in the richest clusters of galaxies.  The galaxy number 
counts drop sharply around the characteristic luminosity
($L_*$)\footnote{In this paper, we use $M^*$ (the characteristic
  absolute magnitude of the knee of the luminosity function) 
  and $L_*$ (the corresponding luminosity) interchangeably.} so 
fitting functions used to describe the luminosity function (expressed in magnitude) have a decaying 
bright end -- \cite{Abe65} used a power-law, \cite{Hub36} a Gaussian, while
\cite{Sch76} a double exponential -- all of which by construction
predict a small number at the brightest extreme.  Hence, the
occurrence of even a single galaxy at large luminosity ($\ga
10\,L_*$) would seem improbable. However, early studies of composite
cluster luminosity functions show an ubiquitous bright end ``hump''
populated by the brightest members. For example, in his seminal paper,
\cite{Sch76} found that the Gamma distribution (in luminosity) provides an excellent fit to the
composite galaxy luminosity function of $13$ massive clusters of
galaxies in the local universe when the brightest galaxy of each
cluster is excluded from the fit. 
This result has been reproduced by
many investigators \citep{Col89,Lug89,Val97,Lum97,Tren98,Gar99,Pao01,Yag02,Got03}.

With the advent of wide-angle redshift surveys, the field galaxy
luminosity function has been increasingly well measured
\citep{Lov92,Nor02,Bla03}. In these data, the bright hump seen in the
galaxy luminosity function derived from rich clusters is either absent or not
as pronounce.
When careful correction for \cite{Edd31} bias due to photometric
scatter is included in the likelihood fit to the luminosity function,
the extreme plunging in number counts predicted from a
double exponential, is confirmed both locally \citep{Bla03} and up to $z
\sim 0.4$ \citep{Loh03}.
This prompted \cite{Sch02} among others to conclude that
the bright end shape of the galaxy luminosity function is universal-- it 
cuts off exponentially in luminosity,
matching the analytical mass function in the \cite{PS74} formalism --
lending credence to the mass upper bound calculations of \cite{Ree77}.

In addition to their extreme luminosities, brightest galaxies in clusters are
often distinguished as having disturbed morphologies and extended
low surface brightness stellar halos. 
Ostriker and collaborators (\citealt{Ost75,Ost77,Hau78},
but see also \citealt{Ric75,Ric76,Whi76,Bin77a,Gun77})
have proposed that these properties are a
consequence of being centrally located in high density environments.
Indeed, the majority of such objects are found in high density environments, and
only rarely does the luminosity function of
moderately rich groups \citep[cf.][]{Gel83}
exhibit a hump.
Hence, accretion via tidal stripping, and merger activity via
dynamical frinction in rich environments would seem to be responsible 
for their growth. 
However, careful numerical and analytical modeling of cluster environments 
\citep{Mer84,Mer85,Tre90} suggest that the luminosities of Brightest
Cluster Galaxies (BCGs) increase only slightly in a Hubble time, as the large
velocity dispersions of the galaxy in these rich clusters makes the
proposed cannibalistic scenario inefficient. Cannibalism, if
it were to operate solely at the present epoch, is insufficient to explain the observed
magnitude difference $\Delta m_{12}$ (hereafter the {\it dominance})
between the BCG and the second brightest galaxy in rich clusters. 
Recent state of the art numerical simulations of massive clusters in a cosmological
context (\citealt{Gao03,Ath01,Dub98}; see also earlier work by \citealt{Wes94,Bod94})
suggest that BCG dominance is driven primary by cosmological infall,
and it thus develops as the clusters themselves form. 

In this paper, we use data from the Sloan Digital Sky Survey (SDSS;
\citealt{Yor00}) to examine the dependence of the luminosity difference
between the BCG and the second-ranked member on environment and
redshift, to test these various evolutionary scenarios.  
 Rather than selecting galaxies in {\it a priori} chosen dense 
environments, we sample the field for the most luminous red galaxies
(i.e., those with the largest stellar masses) at each 
cosmological epoch, using a selection algorithm based on colors and apparent 
magnitudes of a passively evolving old stellar population.
  
Specifically, this paper investigates the bright end behavior of the
luminosity function in regions that host at least one spectroscopic
Luminous Red Galaxy \citep[hereafter LRG;][]{Eis01} from the SDSS.
All high-density regions contain one or more LRGs but LRGs also sample 
lower density environments \citep[e.g.][]{Loh05b,Eis04}, 
including objects without companions to the flux limit of the SDSS
within $\sim 1.0\,\hmpc$.
This work complements older studies using samples selected by 
galaxy overdensities \citep{Tre77,Gel83,Bha85} and 
can test the dominance-overdensity hypothesis. 
It also has a potentially cleaner interpretation, as 
overdensity-selected samples are often subjected to bias from projection effects, and 
by the so-called \emph{selection} effect \citep{Sco57} when comparing samples over a 
range of redshift. 

The outline of the paper is as follows: In the next section, we
describe the sample used in our analysis. We describe the 
statistical tests used in section~\ref{sec:statistics}.  In
section~\ref{sec:gap_analysis}, we outline  
in detail the implementation of our analysis, including a description
of the various subsamples used. In section \ref{sec:gap_results}, we
present our results and give caveats that need to be taken into
account for robust interpretation in
section~\ref{sec:discussion}. Finally, in
section~\ref{sec:gap_conclude}, we summarize our key
conclusions. Throughout 
this paper, $h=H_0/100\kmsmpc$ and we adopt the concordance $\om =
0.3, \ol = 0.7 $ cosmology.

\section{Sample}\label{sec:gap_sample}
\subsection{The SDSS Survey} 
The SDSS will 
eventually image a quarter of the Celestial Sphere and obtain spectra
of approximately 1 million sources. The 5~band (\emph{ugriz},
\citealt{Fug96}) imaging is done  
simultaneously during photometric conditions \citep{Hog01} using a specially 
designed wide-field camera \citep{Gun98} in drift-scanning mode. The imaging data 
are processed by automated pipelines that detect and measure photometric properties 
\citep{Lup01}, and astrometrically and photometrically 
calibrated the data \citep{Pie03,Smi02,Ive04}. From the 
imaging survey, sources are selected for spectroscopy using a 
640 fiber spectrograph mounted on the same telescope. To date, SDSS
has had four major public data releases:  
the Early Data Release \citep[EDR;][]{Sto02}, Data Release One \citep[DR1;][]{Aba03}, 
Data Release Two \citep[DR2;][]{Aba04} and Data Release Three \citep[DR3;][]{Aba05}. 
The data used for this paper are drawn from DR1. 

\subsection{Luminous Red Galaxy Sample}
The LRG sample \citep{Eis01} is a set of intrinsically red and luminous
($\ga\!3L^*$) galaxies targeted spectroscopically by SDSS to create 
a large-volume galaxy survey at moderate number density, out to 
$z\approx0.5$. The full LRG sample comprises $\sim 12\%$ of all SDSS 
galaxy spectroscopic targets \citep{Str02}.  Here, we use those LRGs 
selected by ``Cut I'' -- galaxies that lie on the linear locus of 
$g-r$ and $r-i$ color space -- to give an approximately 
volume-limited sample to $r_{\rm lim} = 19.2$ at $z <0.38$.  

The luminosities and colors of giant elliptical galaxies are observed to 
evolve slowly, and aside from the outer stellar envelope of BCGs and cDs, 
the stars in these galaxies are believed to have formed at high redshift,
e.g. $z > 2$ \citep{Gun75,Ell97,Ara98,Sta98,van98,Bur00}.  The LRG
target selection, described in detail by \cite{Eis01}, 
tunes its selection criteria (based on the uniform SDSS photometry) to
match the luminosity and color of a passively evolving old stellar
population.  The K-corrections as a function of redshift are such that
the contamination from intrinsically blue and low-luminosity objects
is essentially negligible for $0.23 < z < 0.38$. 
Morphologically, the LRG are bulge-dominated galaxies, 
and have surface brightness distributions and stellar light
concentrations similar to those present day giant ellipticals and
lenticulars.  The LRG spectra match the spectral energy distribution
of an old stellar population, although there are unresolved issues 
regarding non-solar abundance ratios \citep{Eis03}.

However, the LRG target selection algorithm only yields a consistent 
population of galaxies for $z > 0.23$. At lower redshift, the color
criteria do not distinguish between less luminous and intrinsically
bluer galaxies from the desired luminous and red ellipticals.  
However, luminous red ellipticals with $z <0.23$ are bright enough 
to be included in the main SDSS galaxy sample \citep{Str02} of which 
rest frame colors and luminosity to select a consistent population.  We use 
an empirically determined color-redshift relation, described
in Appendix \ref{trimmed_LRG} And outlined in full in \cite{Loh05}, 
to select $z < 0.23$ galaxies with the same evolution-compensated rest-frame 
color and luminosities as the higher-redshift LRGs. 

\begin{figure}
\includegraphics[width=0.48\textwidth]{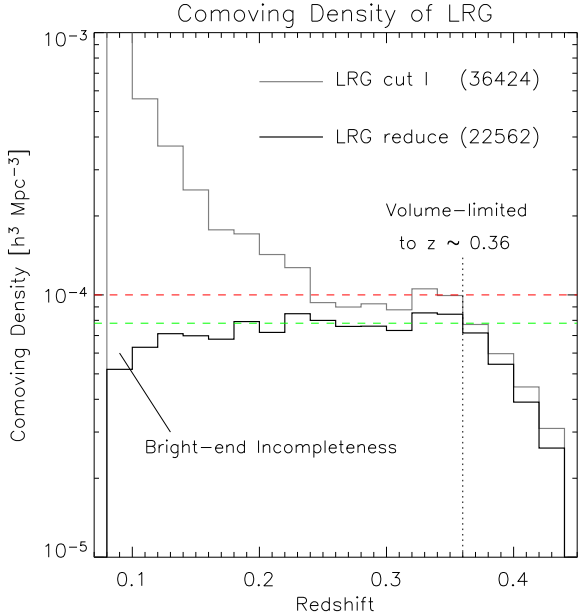}
\caption{
The comoving density of LRG (Cut I) as a function of redshift before (grey) and after (black) 
applying the consistent rest-frame color and luminosity cut as described in Appendix \ref{trimmed_LRG}. 
The new trimmed LRG sample is approximately volume-limited
to $z \approx 0.38$ with a density of $\sim 8 \times 10^{-5} \ihmpcC$. 
\label{fig:lrg_hist}}
\end{figure}

Figure \ref{fig:lrg_hist} shows the histogram of the comoving density of 
LRG before and after we do the consistent rest-frame selection.  
Our sample has essentially a constant density, and is thus volume-limited 
from $z \approx 0.12$ up to roughly  $z \approx 0.38$. The low redshift limit 
is a consequence of the bright-end incompleteness of the SDSS spectroscopic 
survey, as the SDSS spectrographs show excessive crosstalk when the flux within 
the $3 \arcsec$ fiber  exceeds $r\sim 15$ (or $\rpet \sim 13.5$). The high redshift limit comes from the
LRG faint magnitude limit of $r_{\rm lim} =19.2$.
The comoving density of these galaxies is $\sim 8 \times 10^{-5}
\ihmpcC$, an order of magnitude larger than the abundance of 
\cite{Abe65} Richness Class $\geq 0$ clusters of galaxies, and roughly
matches the expected abundance of haloes with mass greater than 
$5 \times 10^{13}\,h^{-1}\,M_{\sun}$ given a concordance power spectrum. 

Our primary spectroscopic data are derived from the uniform
large-scale structure compilation {\tt sample12} of \cite{Bla04};
this covers $2099\,\sqdeg$ of sky.  A thorough discussion of this sample is 
given in the appendix of \cite{Teg04} and \cite{Bla04}.  
There are $12,608$ LRG with $0.12 < z < 0.38$ in {\tt sample12} that pass the
consistent population rest-frame magnitude and color cuts 
described in Appendix \ref{trimmed_LRG}. We refer to this as the
\emph{trimmed} LRG sample.  To determine the environments of the LRG
in this sample, and to measure the dominance of the BCG, we use imaging 
data from the SDSS DR1.  We consider 
all sources that are classified as galaxies by the SDSS photometric
pipeline \photo with $\rpet < 21.0$, a level at which the photometry
is of very high S/N, and at which star-galaxy separation is
particularly clean \citep{Scr02}. $\rpet$, the \cite{Pet76}
magnitude in the SDSS $r$-bandpass, is the primary extended source
flux measurement used in this paper; see \cite{Str02} for a complete
discussion.  All magnitudes and colors are corrected for Galactic extinction
using the \citet{Sch98} map and assuming $R_V=3.1$.  

\section{Statistical Tests of the Luminosity Function Tail} 
\label{sec:statistics}
The observed narrow luminosity and color distribution of BCGs
\citep{Pos95}, and the possible influence of dynamical friction and
tidal stripping in their evolution, suggest that their luminosity
function may be distinct from that of the field. 
This is often framed in terms of \emph{statistical} versus \emph{special} 
luminosity functions: the former argues that BCGs are a mere 
statistical extreme of the luminosity function of cluster ellipticals, 
while in the latter, the BCG forms a different class, having its own unique
distribution shared among BCGs in many clusters.

\cite{Tre77} have developed an elegant test to distinguish between
these two descriptions; this method does not 
require the determination of the galaxy luminosity functions
of the individual clusters involved. Readers who are interested in the detailed
derivation of the theorems that support the validity of the test
should refer to their paper. The test hinges on two key assumptions
about the nature of the galaxy luminosity distribution in clusters.
The first is \cite{Sco57}'s model for the luminosity function, which 
postulates that for a given luminosity distribution, overlapping
magnitude intervals are statistically independent.  If $\psi$ is the
integrated luminosity function 
\beqa 
\psi(m) &=& \int_{-\infty}^{m}
\phi(m')\,\ud m'\ , 
\eeqa 
then the probability of finding $\nu$ galaxies
in the magnitude interval [$m_a,m_b$] is 
\beqa 
p_{\nu}(m_a,m_b) &=&
\frac{\left[\psi(m_b)-\psi(m_a)\right]^{\,\nu}}{\nu\,!}
\exp\left[\psi(m_a)-\psi(m_b)\right] \label{eq:sm1}. 
  \eeqa 
The number, $\nu$, of galaxies
(within a single isolated cluster) is a \textit{Poisson} variate
with mean and variance $\psi(m_b) - \psi(m_a)$.  In the case of finding a single
galaxy in the infinitesimal interval [$m,m+\ud m$],
equation~(\ref{eq:sm1}) reduces to 
\beqa 
\ud P &=& p_1(m,m+\ud m)
\,\,\, = 
\,\,\,\phi(m)\,\ud m 
\eeqa 
In this model, the probability distribution
of the $j$-th ranked galaxy with magnitude $m$, $p_{(j)}(m) \ud m$, is
just\footnote{\normalsize{Numbers in parentheses, e.g. $m_{(j)}$,
    indicate ($j$-th) ranked variables.}}  
\beqa
p_{(j)}(m)\,\ud m &=& {\rm prob}\,\{j-1\,\,\textrm{galaxies brighter than}\,m\}\nonumber\\
& & \times\,{\rm prob}\,\{\textrm{one galaxy in}\,(m,m+\ud m)\}\nonumber\\
&=& p_{j-1}(-\infty,m) \times p_1(m,m+\ud m)\nonumber\\
&=&
\frac{\psi(m)^{j-1}}{(j-1)\,!}\exp[-\psi(m)]\,\phi(m)\,\ud m
\label{eq:poisamp} 
\eeqa 
\citeauthor{Sco57}'s model thus states that galaxy luminosities are
drawn from a Poisson distribution, with magnitudes $m$ being
treated as independent random variables. This is completely embodied
in equation (\ref{eq:poisamp}).

The second assumption is that the bright end of the
luminosity function drops off as a
power-law (exponentially in magnitude), which allows one to write down 
an explicit analytical expression for the luminosity distribution of
the $n$th-ranked galaxy.   For example, the probability distribution of the
luminosity of the first-ranked galaxy with a universal bright end
differential luminosity function $\phi(m) \simeq \exp[\alpha(m -
m_0)]$ is \beqa
p_{(1)}(m)\,\ud m &=& \exp[-\psi(m)]\,\phi(m)\,\ud m\nonumber\\
&\simeq& \alpha\exp[\alpha(m - m_0) - e^{\alpha(m - m_0)}]\,\ud m.
\label{eq:first_rank}
\eeqa Here, $\alpha$ parameterizes the steepness of the function, and
$m_0$ gives the normalization (or richness) of the cluster. 

Early studies by \cite{Pee68,Pet70,Gel76,Sch76b} 
that compared observed ranked magnitude distributions with 
\citeauthor{Sco57}'s model prediction (equation
(\ref{eq:first_rank}) and its extension to higher ranks) were
inconclusive. This is because the catalogs used in these studies had
insufficient redshift information and reliable cluster membership for
even the brightest few galaxies. In addition, small uncertainties in
the relative \emph{k}-correction in clusters at various distances wash
away subtle effects that would otherwise distinguish the statistical
and special descriptions of the BCG luminosity distribution.

\subsection{The Magnitude Difference Between First and Second-Ranked Galaxies}
\citeauthor{Tre77}'s key insight was to realize the importance of
using the magnitude differences for a statistical test. The
exponential cutoff in the differential luminosity function $\phi
\sim \exp(\alpha m)$ not only predicts the expected distribution of
the first-ranked magnitudes (equation \ref{eq:first_rank}), but also
puts a tight constraint on the expected distribution of the
\emph{differences} between the ranked magnitudes.
BCGs (the first-ranked members) are observed to have a small spread in 
magnitude, $\sigma(M_1) < 0.3$ \citep{San72,Pos95}, suggesting
that on average, the magnitude difference between the first-ranked galaxy,
and their respective second-ranked galaxies, $M_{12} \equiv M_2 -
M_1$, cannot be too large, if both of these galaxies are drawn from 
the same exponentially decaying $\phi$.
Indeed, \citeauthor{Tre77} showed that the expected magnitude
difference, $\langle M_{12} \rangle$ must at most be of the same order
as the size of the spread of the first ranked galaxy magnitudes: 
\beqa 
T_1 &\equiv&
\frac{\sigma(M_1)}{\langle M_{12}\rangle}~\geq~1\ . \label{eq:t1} 
\eeqa
Similarly, they showed that the spread of the magnitude
difference, $\sigma(M_{12})$, must also be of the same order of
magnitude as the difference $\langle M_{12} \rangle$: 
\beqa 
T_2 &\equiv& \frac{\sigma(M_{12})}{\langle
  M_{12}\rangle}~\ga~0.82 \ .
\label{eq:t2}
\eeqa
The two inequalities are valid even when there are variations in the
bright end slope $\alpha_i$ of each of the clusters of galaxies used
to derive the distributions of $M_1$, $M_2$, $M_{12}$ etc., so long as
the $\phi_i$ satisfy the two assumptions discussed above. Note that  
equation (\ref{eq:first_rank}) holds only under the more restrictive
assumptions that all $\alpha$ take a single value, and that the 
intrinsic richness of the clusters, $m_0$ be the same.

However, in the limit that all the $\alpha_i \rightarrow \langle
\alpha \rangle < 0$, the magnitudes, $m$, of \citeauthor{Sco57}'s model
would be an identical and independently distributed (i.i.d) random
variable. Using extreme-value statistics, \cite{Bha85} showed that as
long as $\phi$ has a general functional form in the \emph{domain of
  attraction} of the exponential distribution -- including
the Gaussian, Gamma, exponential and double-exponential distributions,
then equation (\ref{eq:first_rank}) is the expected distribution for statistical BCG, 
and the expectation value for our two statistics are 
$\langle T_1 \rangle \simeq 1.28$ and $\langle T_2 \rangle \simeq 1.0$.  
It is remarkable that both these values are independent of both $\alpha$ 
and $m_0$.

Note that the statistic $T_2$ depends solely on the differential
quantity $M_{12}$, and eliminates uncertainties associated with
redshift and direction dependent corrections like \emph{k+e} and
extinction, making analyses that combine data from different epochs
more robust.

\section{Measurement of $M_1$, $M_2$ and Richness of LRG Environments}\label{sec:gap_analysis}
The trimmed LRG spectroscopic sample we are using extends to $z =
0.38$.  However, the main SDSS galaxy sample, probing lower-redshift
galaxies, has a median redshift of only $\sim 0.1$, with very few
galaxies above $z = 0.2$.  We therefore use the \emph{photometric}
data from the SDSS to characterize the environments and the luminosity
distribution of galaxies around each LRG, and  to calculate statistics 
such as $M_{12}$. We do so by using color cuts to isolate early-type galaxies 
at the same redshift as the LRG, using the red sequence seen in
cluster color-magnitude diagrams. Within an angular extent equivalent to 
$1.0\,\hmpc$ at the redshift of each spectroscopic LRG, we searched the 
SDSS photometric data for second- and third-ranked galaxies with $g-r$ and
$r-i$ colors typical of early-type galaxies at that redshift. The
regularity of the observed  color-magnitude relation of early-type galaxies 
seen out to $z \sim 0.5$ and beyond, e.g. \cite{Blak03}, enables us to 
empirically determine a color-redshift locus for these galaxies.  This is 
shown in Figure \ref{fig:gr_ri_locus}, with an estimated dispersion at each 
redshift given by the solid ellipses. Details of how this locus is determined is
described in Appendix \ref{appendix}. Note that the locus is not simply 
drawn from the colors of LRG themselves since ``Cut I'' LRGs are red biased 
-- lower luminosity galaxies need to have a redder color to pass the 
selection cut \citep{Eis01}. 
The ellipses include the observed color errors -- almost entirely due to photometric 
scatter\footnote{The intrinsic scatter of the color-magnitude relation 
is expected to be less than 0.05 magnitude even at $z = 0.4$ 
\citep{Ell97,Coo05,Loh05}.} -- of galaxies on the red sequence with 
apparent magnitude brighter than $\sim M^*(z)$ at each redshift. 
Hence, if we include only galaxies from the
fields around each LRG with colors within the $n^{\rm th}\,\sigma$
ellipse of the LRG redshift, then on average, we will be complete with
respect to red sequence galaxies in the LRG fields at the
$n\,\sigma$ level.  Of course, for larger $n$, the background
contamination goes up.  We find that the contamination remains
manageable, while still including the majority of galaxies in the
red sequence, if we include galaxies that are within $2.5
\,\sigma$ of the color locus.  

The galaxies are then ranked by apparent Petrosian $r$
magnitude. The first-ranked galaxy in each 
field is the LRG 95--99\% of the time, with the lower 
values holding in higher redshift fields.  In some of the exceptions,
the first-ranked galaxy was also flagged as an LRG candidate, but a
spectrum was not obtained because of the restriction that two
spectroscopic fibers not be placed closer than $55''$ \citep{Bla03}. 
The majority of the remainder turned out to be bluer galaxies
presumably in the foreground, selected because
of the generosity of the $2.5 \,\sigma$ color cut. 
We keep this small number of non-LRG first-ranked galaxies in our 
analysis so as not to bias our result against the occasional true bluer
first-ranked galaxy. However, excluding them from our sample does
not significantly change our results.

\begin{figure*}
\includegraphics[angle=0,width=0.48\textwidth,height=0.47\textwidth]
{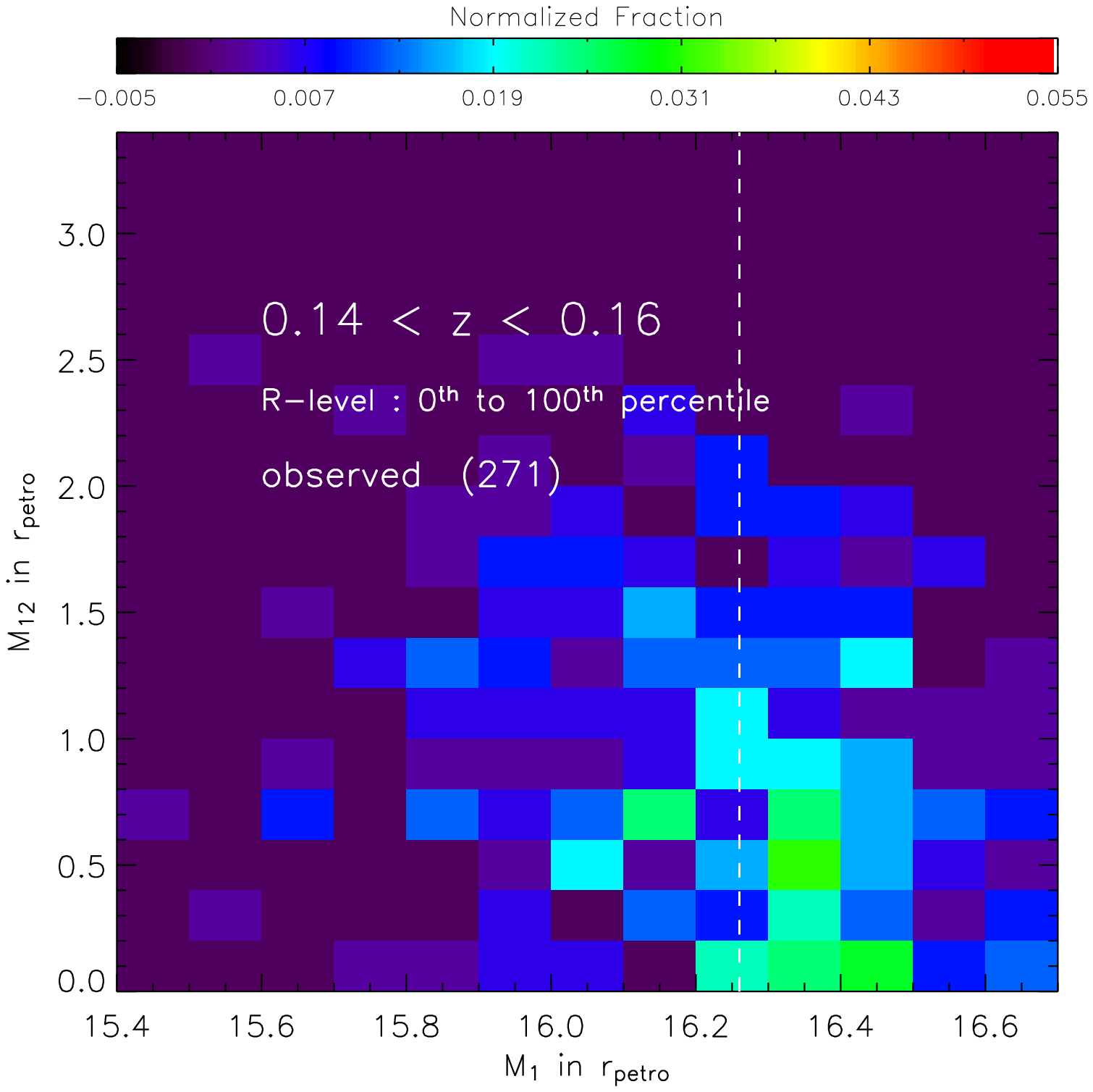}
\includegraphics[angle=0,width=0.48\textwidth,height=0.47\textwidth]
{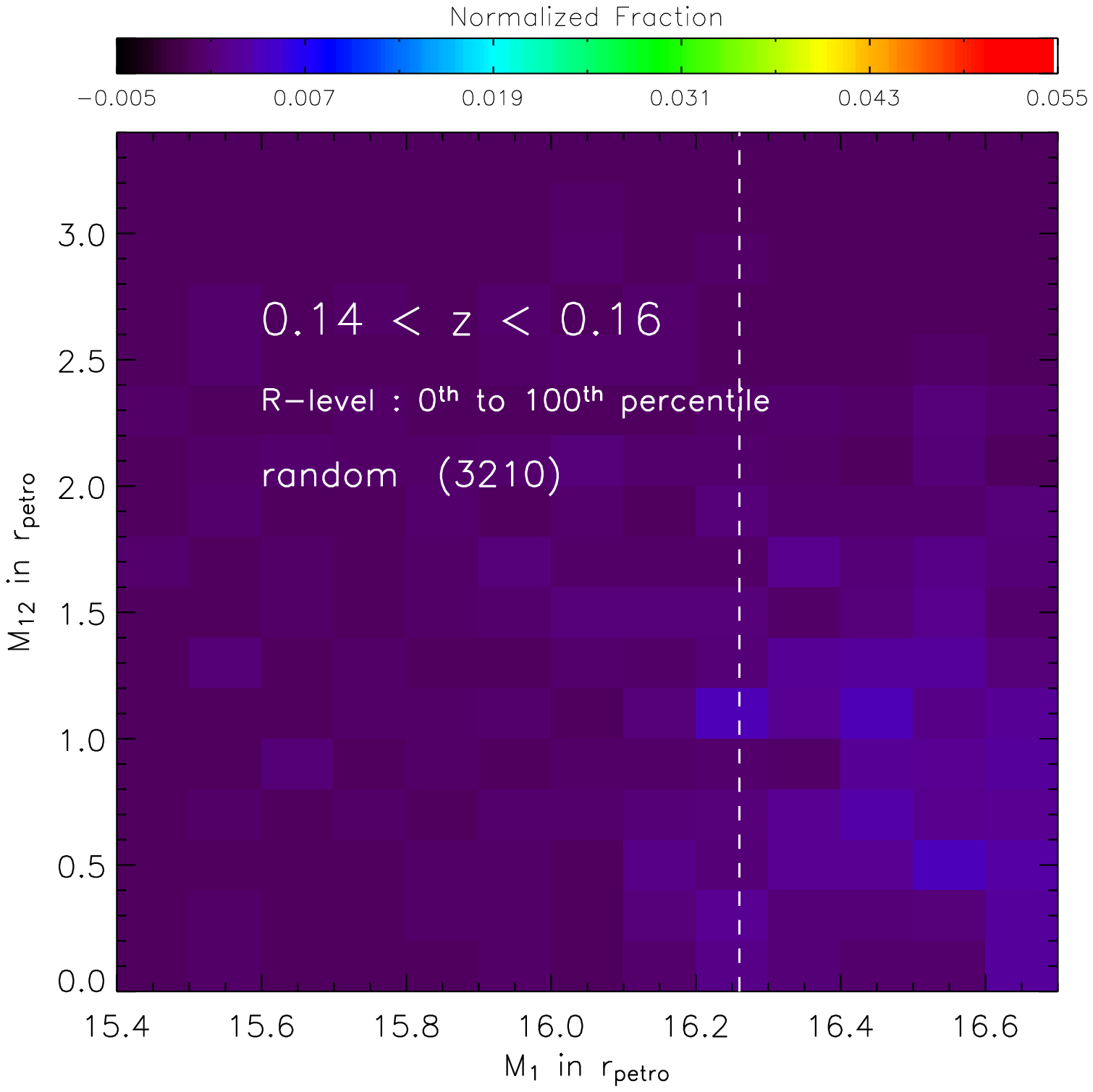}
\includegraphics[angle=0,width=0.48\textwidth,height=0.47\textwidth]
{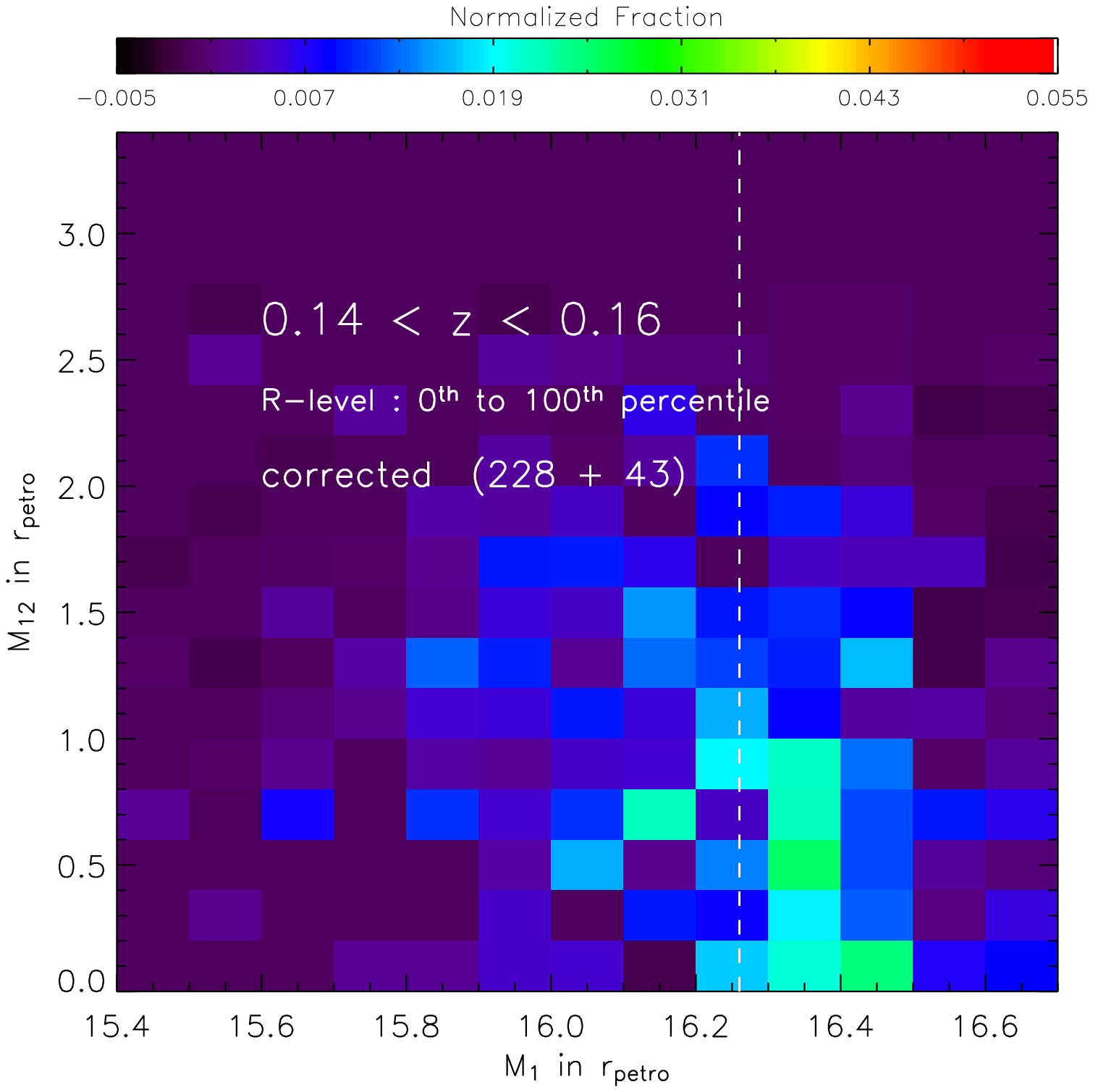}
\includegraphics[angle=0,width=0.48\textwidth,height=0.47\textwidth]
{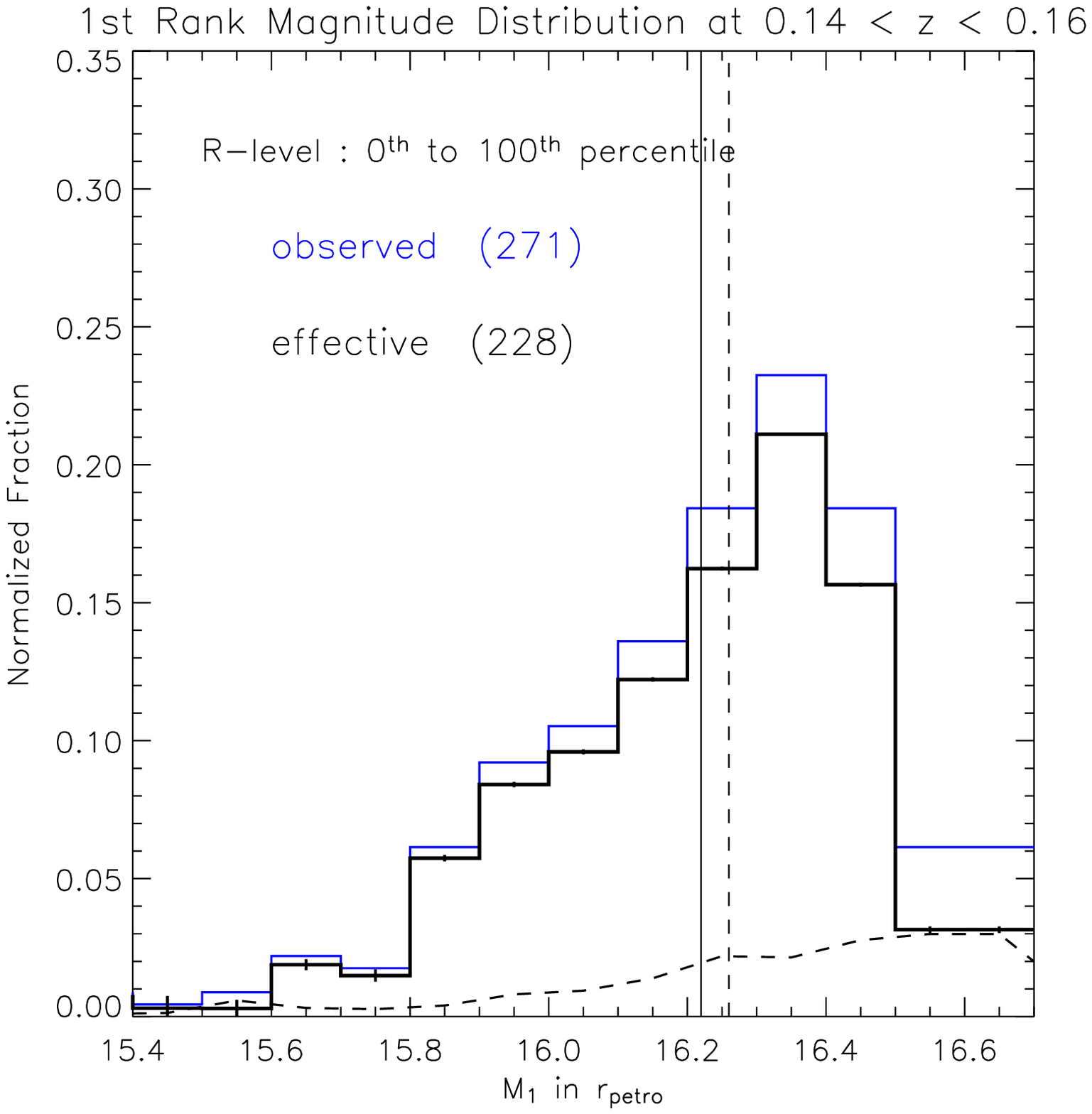}
\caption[]{These figures illustrate the background subtraction in $M_1$ and $M_{12}$
space, done in a narrow redshift interval of $0.14 < z < 0.16$. The top left panel gives 
the observed composite $\phi(M_1,M_{12})$ for 271 LRG fields; top right is the background 
estimated from 3210 random positions, but scaled by area to match the observed $\phi(M_1,M_{12})$.
The bottom left panel is the corrected distribution using equation (\ref{eq:gap_est}), 
weighting by $f = 228/271$, the ratio of the number of uncontaminated (effective) fields 
to the total observed fields. The bottom right gives the marginalized $1^{\rm st}$ ranked 
distribution. The blue histogram is the observed $\phi(M_1)$ distribution, and the black 
histogram is the effective distribution obtained by subtracting the (dashed) background. 
The solid vertical line gives the mean $\langle M_1 \rangle$. The magnitude of a $M^* - 1.5$ 
early-type galaxy at the median redshift of $z \sim 0.15$ is indicated by the dashed vertical 
line in all panels. Density contours and histograms are normalized to unity.
\label{fig:m1_m12_z0.15}}
\end{figure*}
\begin{figure*}
\includegraphics[angle=0, width=0.48\textwidth,height=0.47\textwidth]
{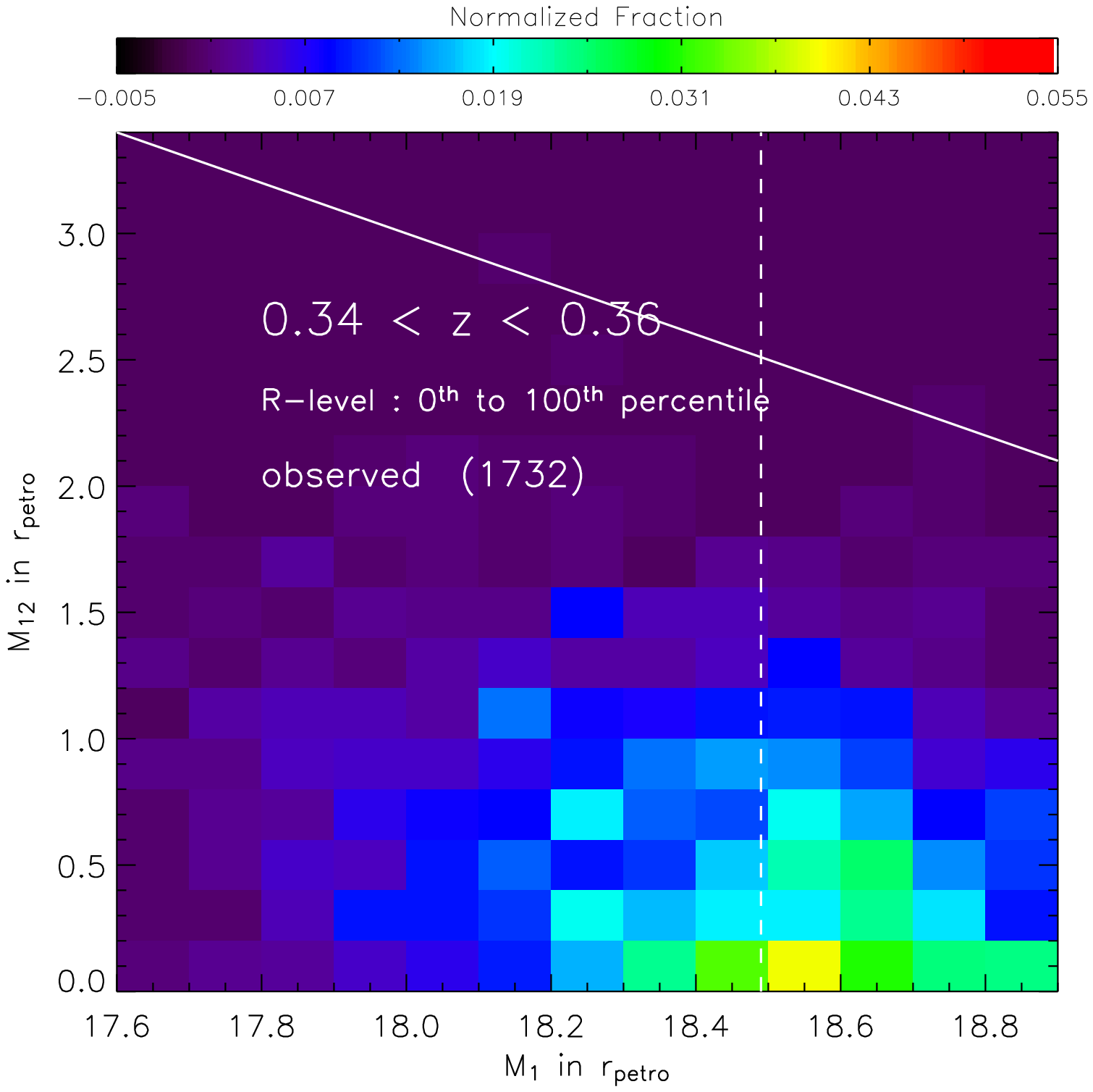}
\includegraphics[angle=0, width=0.48\textwidth,height=0.47\textwidth]
{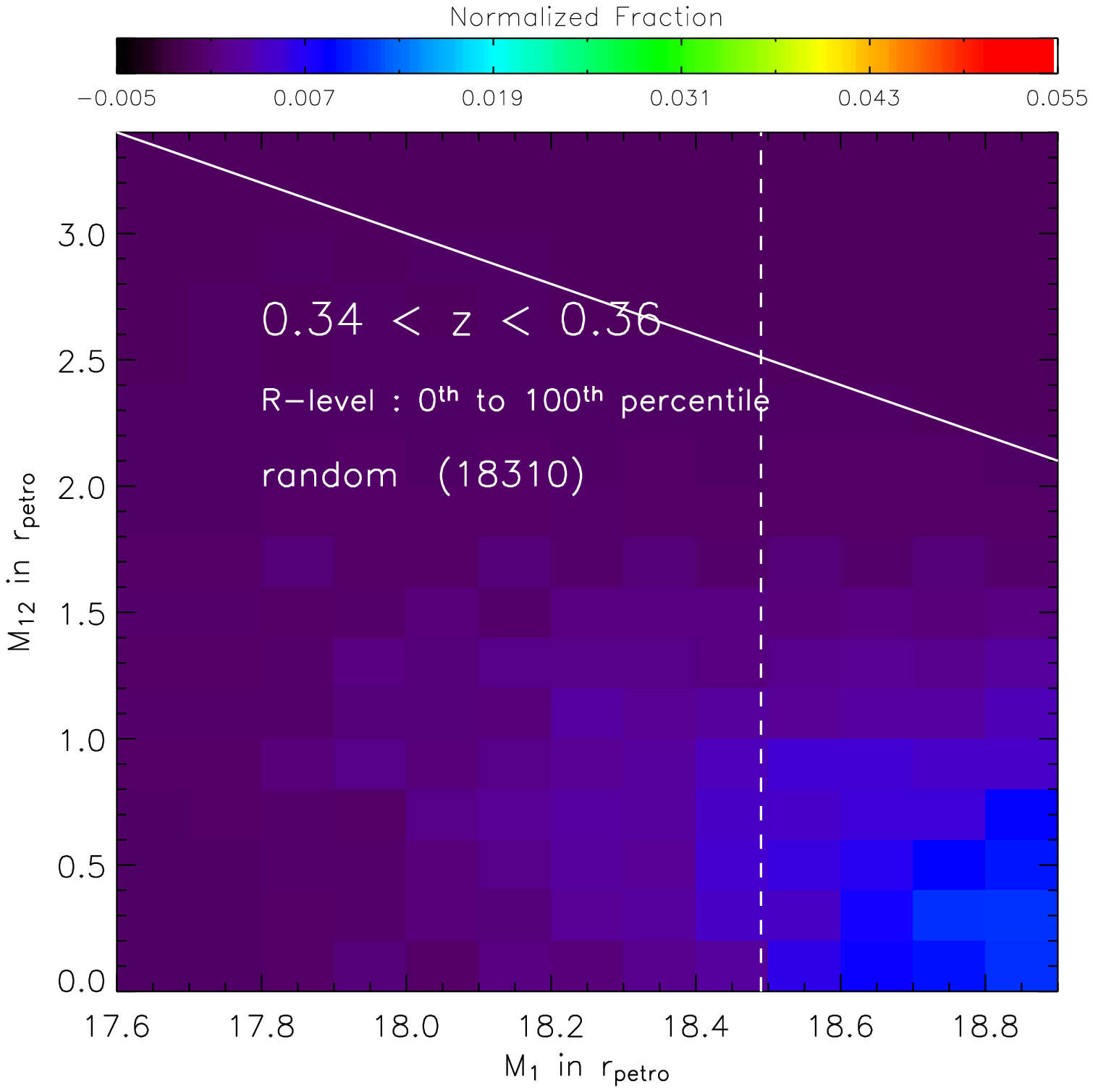}
\includegraphics[angle=0, width=0.48\textwidth,height=0.47\textwidth]
{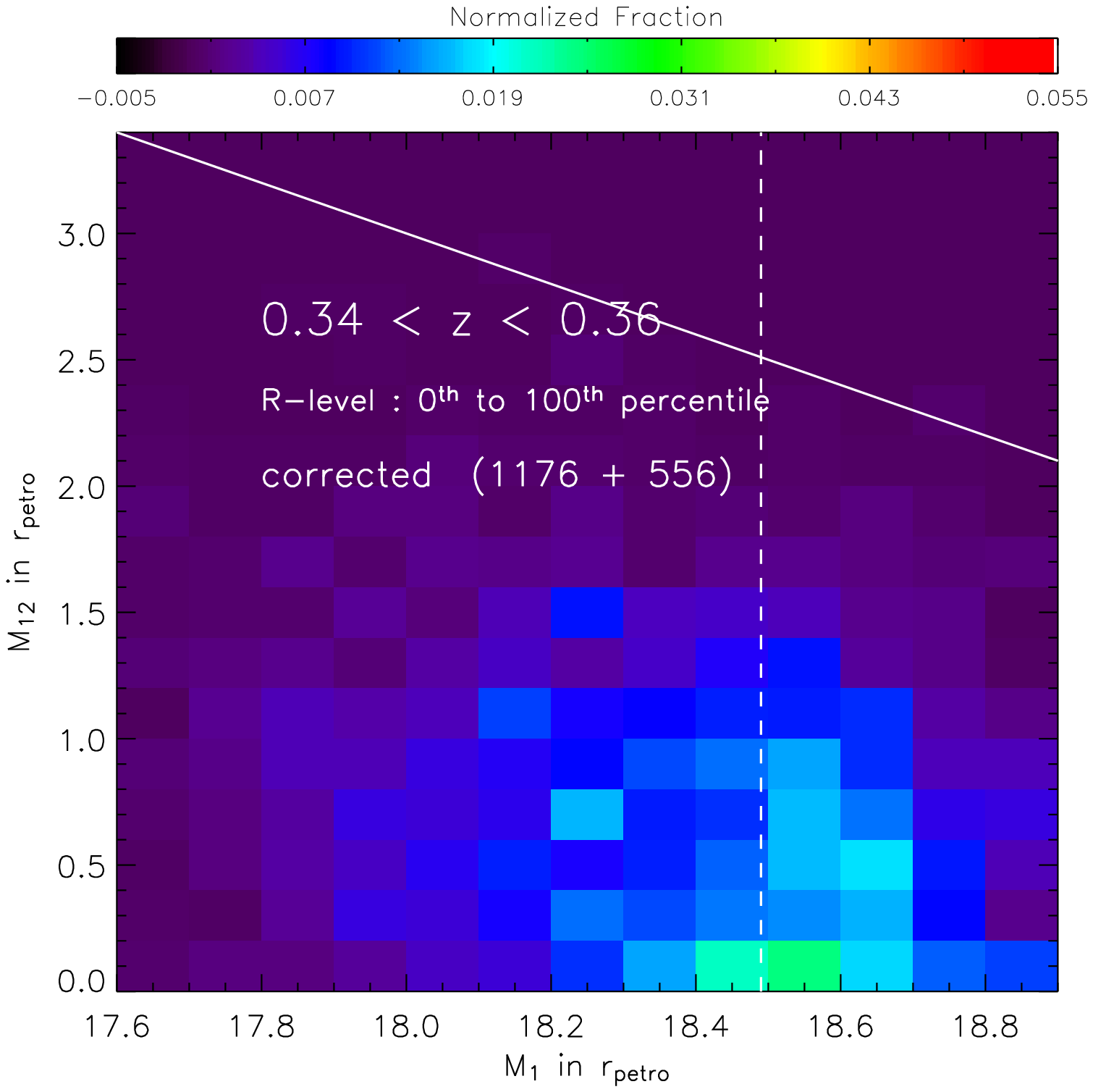}
\includegraphics[angle=0, width=0.48\textwidth,height=0.47\textwidth]
{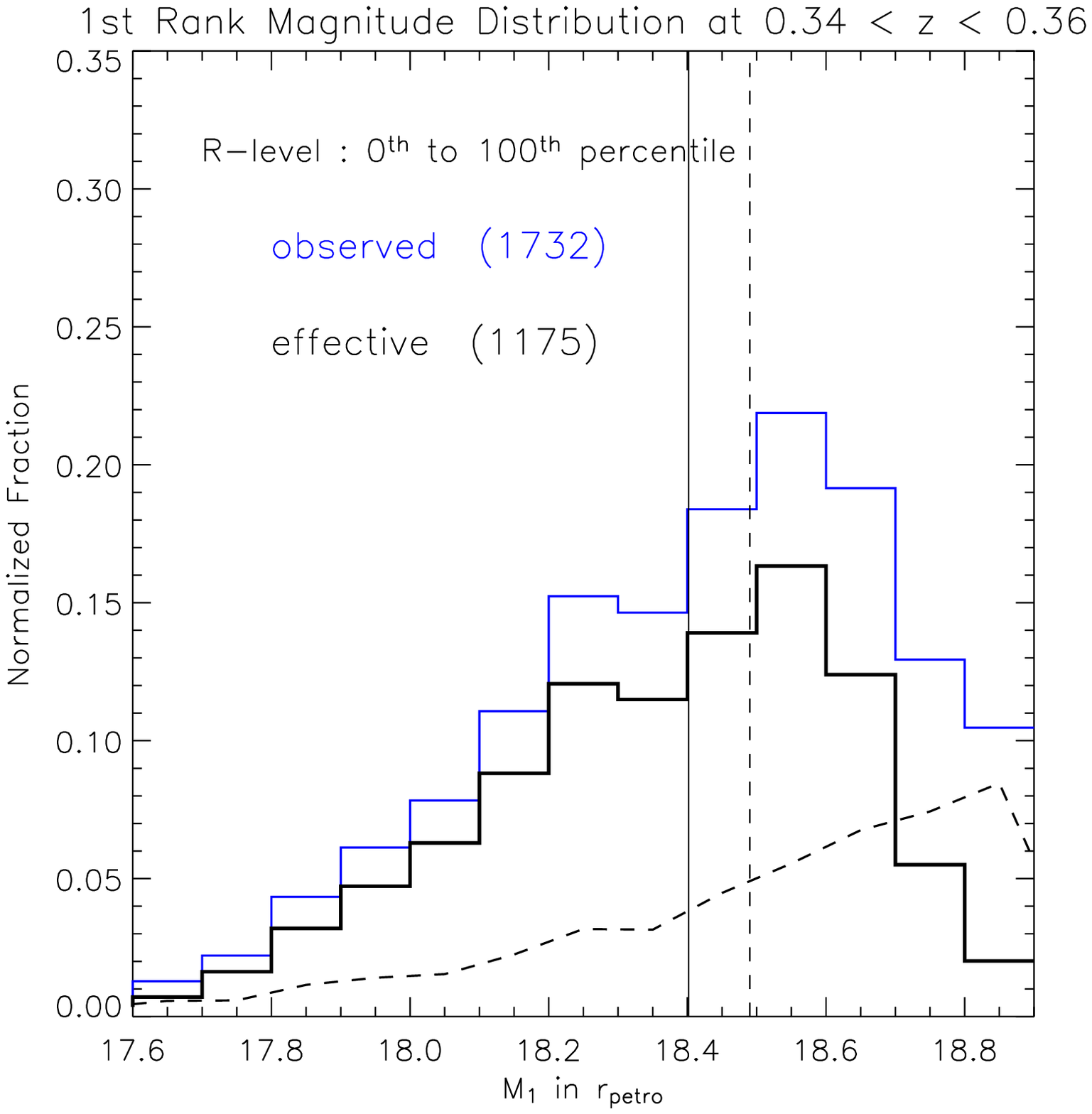}
\caption[]{As in Fig. \ref{fig:m1_m12_z0.15}; for $0.34 < z < 0.36$. The solid 
diagonal line in the 2D plots gives the upper bound of $M_{12}$ for a given $M_1$ set
by the flux limit ($\rpet < 21.0$) of the SDSS imaging catalog. 
\label{fig:m1_m12_z0.35}}
\end{figure*}
Our first goal is to derive the intrinsic joint distribution of 
$M_1$ and $M_{12}$, $\phi(M_1,M_{12})$. For this, we need to estimate 
and subtract the background distribution $\phi^{\rm back}(M_1,M_{12})$. 
For each LRG field, we select ten random positions within the same
SDSS imaging {\tt stripe} \citep{Yor00}.  Just as in the LRG fields,
we draw a $1.0\,\hmpc$ circle, apply the same 2.5\,$\sigma$ red
sequence color cut, and rank the galaxies by brightness.  The $\phi$ are binned
with $\Delta M_1 = 0.1$ {\tt mag} and $\Delta M_{12} = 0.2$ {\tt mag}. 
Figures \ref{fig:m1_m12_z0.15} and \ref{fig:m1_m12_z0.35} show the composite $\phi$ for 271 
and 1732 LRG fields at $0.14 < z < 0.16$, and $0.34 < z < 0.36$, respectively. 
The top left panel of each figure is the observed
distribution while the bottom left gives the background. The dashed
vertical line is the observed magnitude of a $M^* - 1.5$
galaxy at the respective redshift, while the slanted solid line gives
the upper bound to the $M_{12}$ we could observe given the SDSS
photometric catalog limit of $\rpet = 21.0$. 
Between 0.5\% and 1\% of all LRG in our sample are isolated (see 
\citealt{Loh05b} for further discussion), showing no companion within the color 
and magnitude space we searched.  These objects may well be
fossil groups \citep{Pon94}, showing substantial numbers of low-luminosity
companions below our photometric limits.  We leave the study of these
objects to a future paper. 

We estimate the intrinsic distribution in $M_1$ and $M_{12}$ (given in
the top right panels) using a weighted subtraction given by
\beqa
\hat{\phi}\,(M_1, M_{12}) &=& f \times [\phi^{\rm obs}(M_1,M_{12}) - 
                                        \phi^{\rm back}(M_1,M_{12})]\nonumber\\ 
                          && + (1-f)\times\phi^{\rm obs}(M_1,M_{13})\label{eq:gap_est}
\eeqa 
where $f \equiv (N_{\rm obs} - N_{\rm back})/N_{\rm obs}$ is the fraction of fields in which
$M_{12}$ measurements are not affected by contamination, and $N_{\rm obs}$ and 
$N_{\rm back}$ are the number of LRG and background fields in the
$M_1, M_{12}$ bin in question.   If the
background contamination $(1-f)$ is large, there is a substantial
probability that the galaxy we've labelled the second-ranked is
actually a background (or foreground) object, thus what we have labelled
the third-ranked galaxy is in fact $M_2$.  This is the origin of the
second term in equation~(\ref{eq:gap_est}).  

We first perform our analysis on the full volume-limited trimmed
sample of LRG in a series of redshift intervals of width $\Delta z =
0.02$ from $z = 0.12$ to $z = 0.38$.
We define the richness (or the associated galaxy overdensity) of each 
LRG field by counting the number of galaxies on the red sequence within 
$1.0\,\hmpc$ of the LRG from $M^* - 2.5$ down to $M^*$ \citep{Loh05b}.
We then split the sample into four, in terms of quartiles based on the
richness of these LRG fields; the richness ranking is done within each redshift
slice. We term the relative rank of each LRG in its redshift slice 
the $R$-level. The upper quartile sample, with a median of $8$ early-type galaxies 
with magnitude $M^*$ and brighter within a $1.0\,\hmpc$ radius,  
are the equivalent  
of a present day moderately rich cluster of galaxies (e.g. \cite{Abe65} 
Richness Class $> 1$). Objects in the $25^{th}-50^{th}$ percentile, on
the other hand,  with 2 to 3 galaxies on the red sequence, are similar to 
present day groups of galaxies.

\begin{figure}
\includegraphics[width=0.48\textwidth]{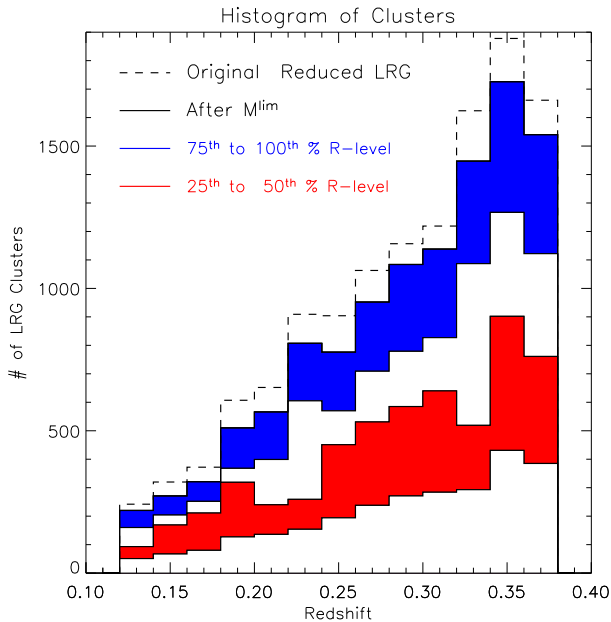}
\caption[]{Redshift histogram of the trimmed LRG sample. 
The blue (red) shading indicate the number of LRG fields with richness in the upper 
$75^{\rm th}$($25^{\rm th}-50^{\rm th}$) percentile. 
\label{fig:hist}}
\end{figure}

Figure \ref{fig:hist} is the redshift histogram of LRG used for our
analysis. The blue shading refers to the upper quartile, while the red shading indicates 
the $25^{th}-50^{th}$ percentile. 
Note that the fraction of LRG fields in the upper
quartile (say) of a given redshift subsample can be greater or smaller
than 25\% of the subsample total because LRGs with the same richness
share the same rank.

\section{Tremaine-Richstone Statistics as Functions of Redshift and
  Richness}\label{sec:gap_results} 

Figures \ref{fig:m1_m12_z0.15} and \ref{fig:m1_m12_z0.35} show that 
as one goes to higher redshift, LRG clusters tend to have smaller $M_{12}$. 
For example, at $z \approx 0.15$, there are quite a few LRG clusters with 
$M_{12} \sim 2$, well above the noise level (given by the bottom left panels), while 
there are no clusters with such a large gap at $z \approx 0.35$.

\begin{figure*}
\includegraphics[angle=0, width=0.48\textwidth]{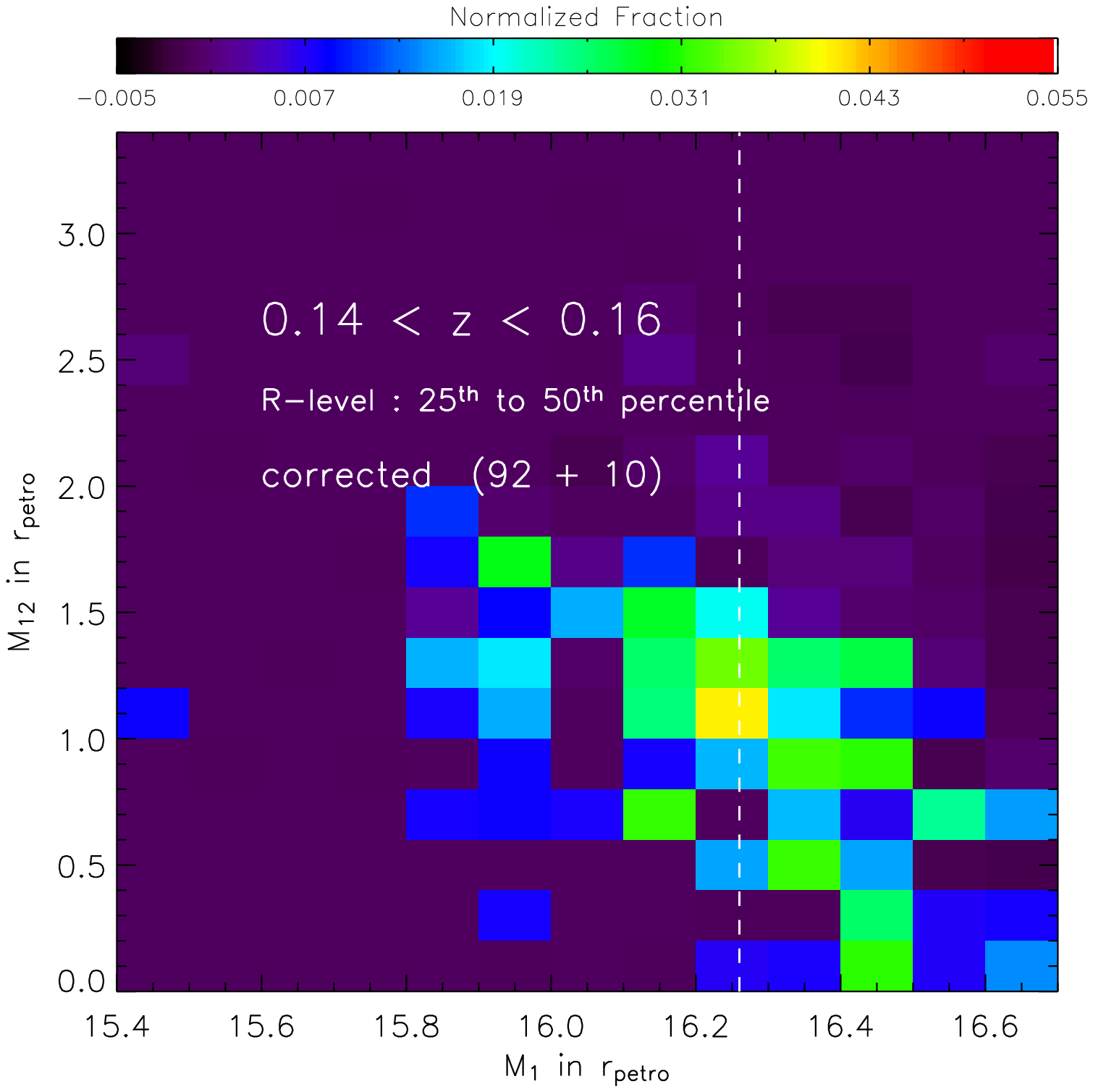}
\includegraphics[angle=0, width=0.48\textwidth]{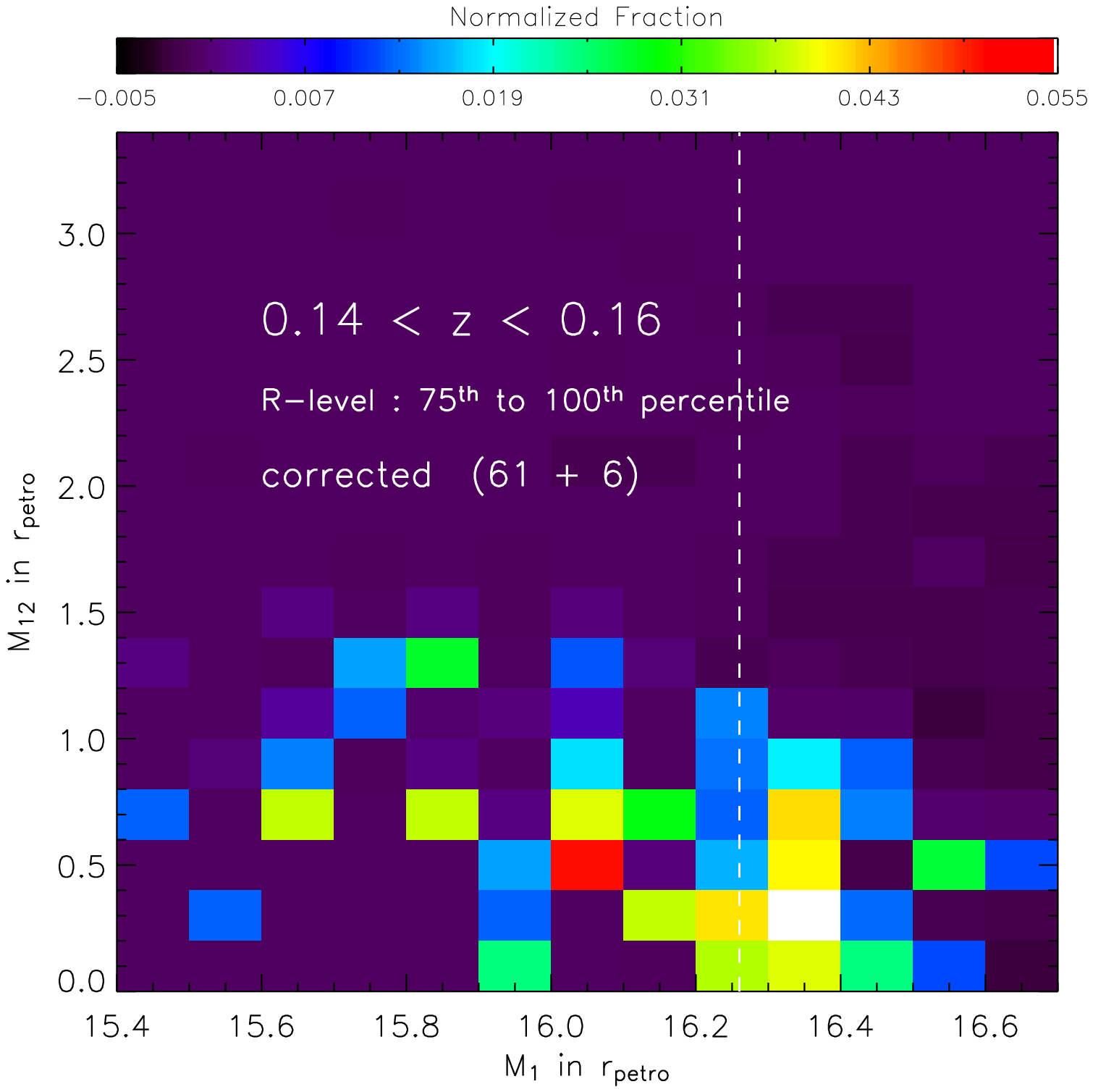}
\includegraphics[angle=0, width=0.48\textwidth]{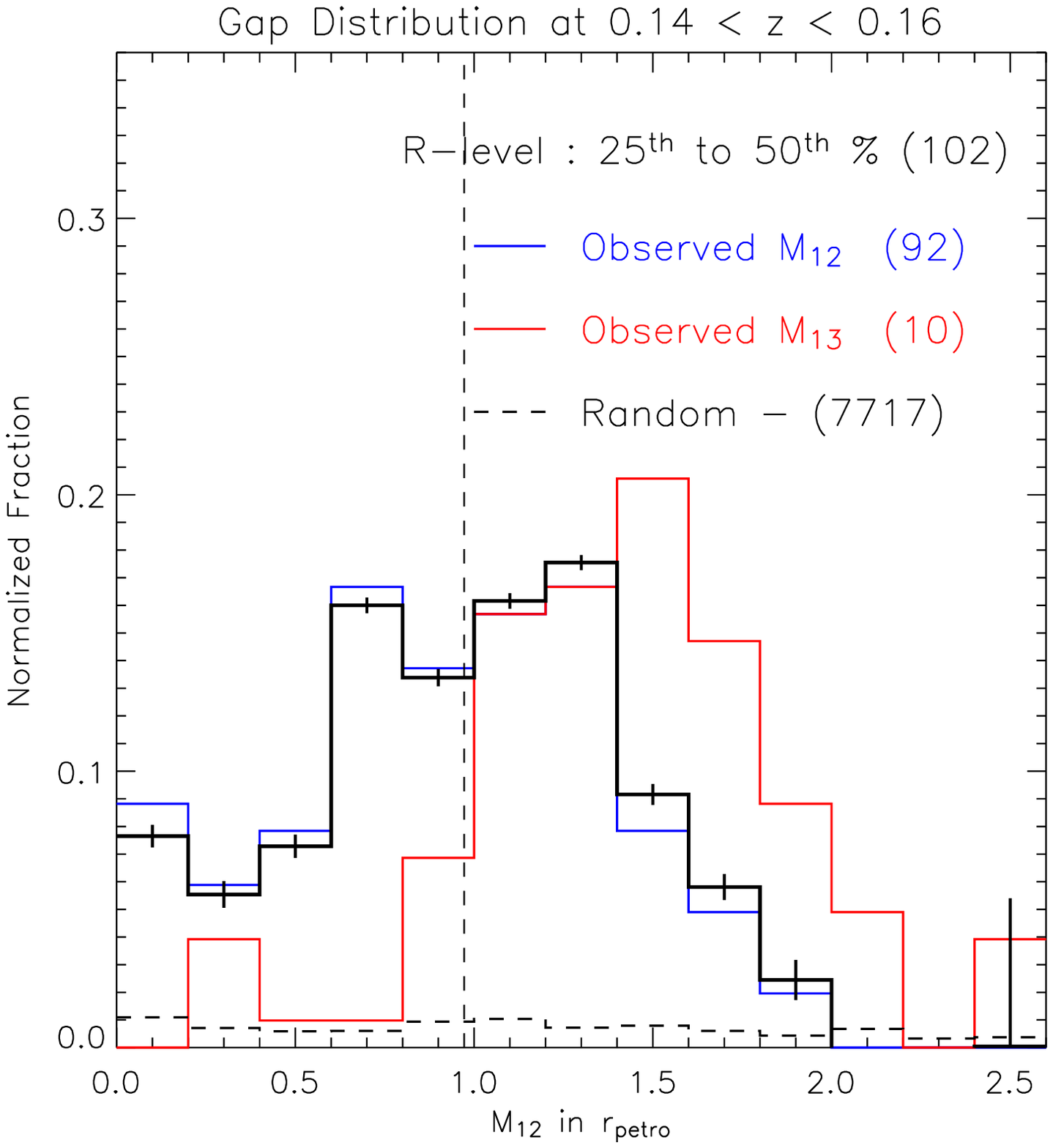}
\includegraphics[angle=0, width=0.48\textwidth]{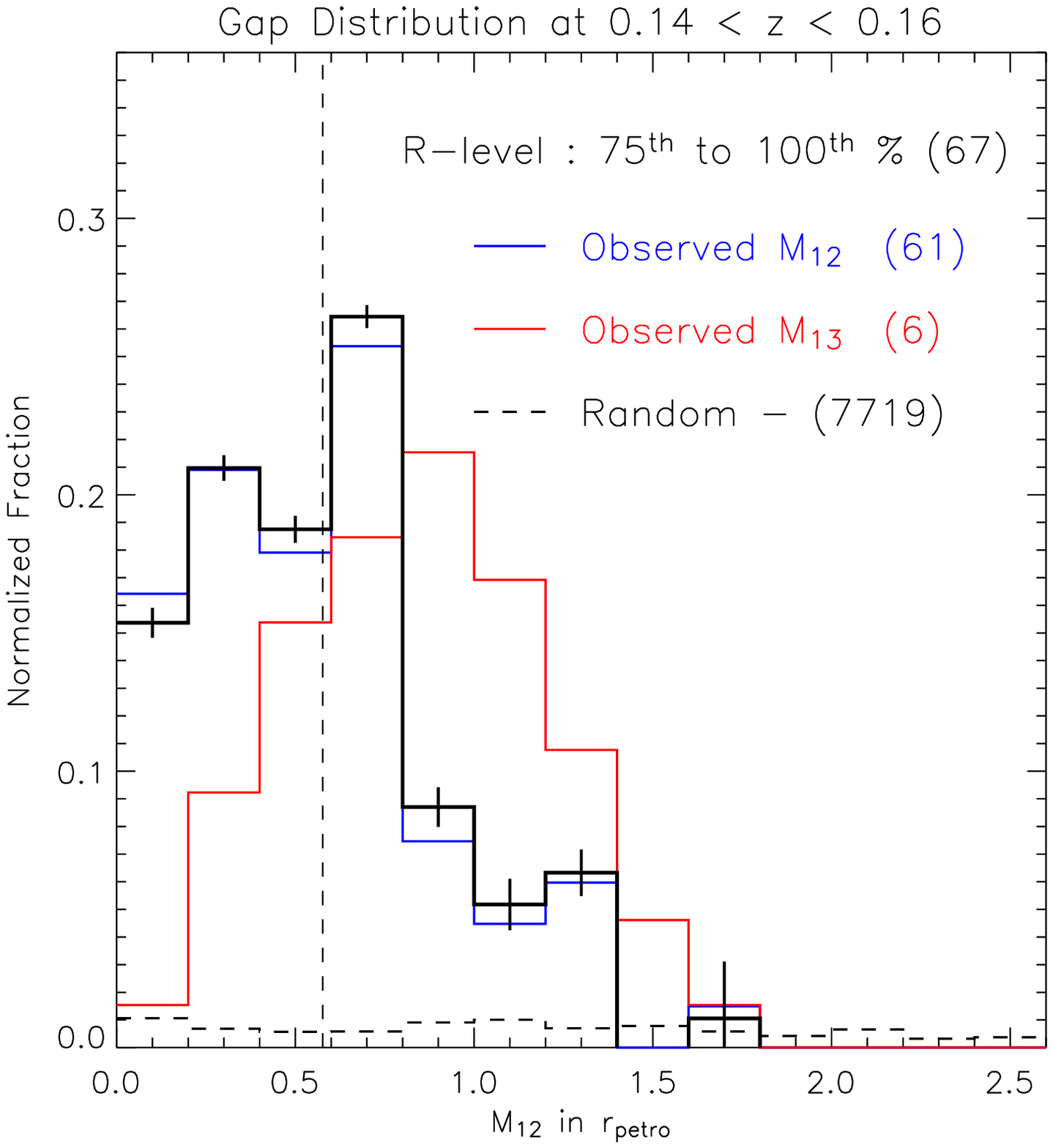}
\caption[]{On the top left (right) panel is $\phi(M_1,M_{12})$ for LRG fields with richness
$R$-level in the $25^{\rm th}$ to $50^{\rm th}$ ($75^{\rm th}$ and above) percentile. 
Higher percentile corresponds to richer systems. The bottom panels
are the corresponding marginalized $\phi(M_{12})$. The black histogram is 
the final distribution estimated from equation (\ref{eq:gap_est}) --
a weighted linear combination of the observed $M_{12}$ (blue) and $(M_{13})$ (red)
distribution, and the background distribution (dashed). For LRG fields 
with redshifts $0.14 < z < 0.16$.  
\label{fig:m12_hist_z0.15}}
\end{figure*}

\begin{figure*}
\noindent
\includegraphics[angle=0, width=0.48\textwidth]{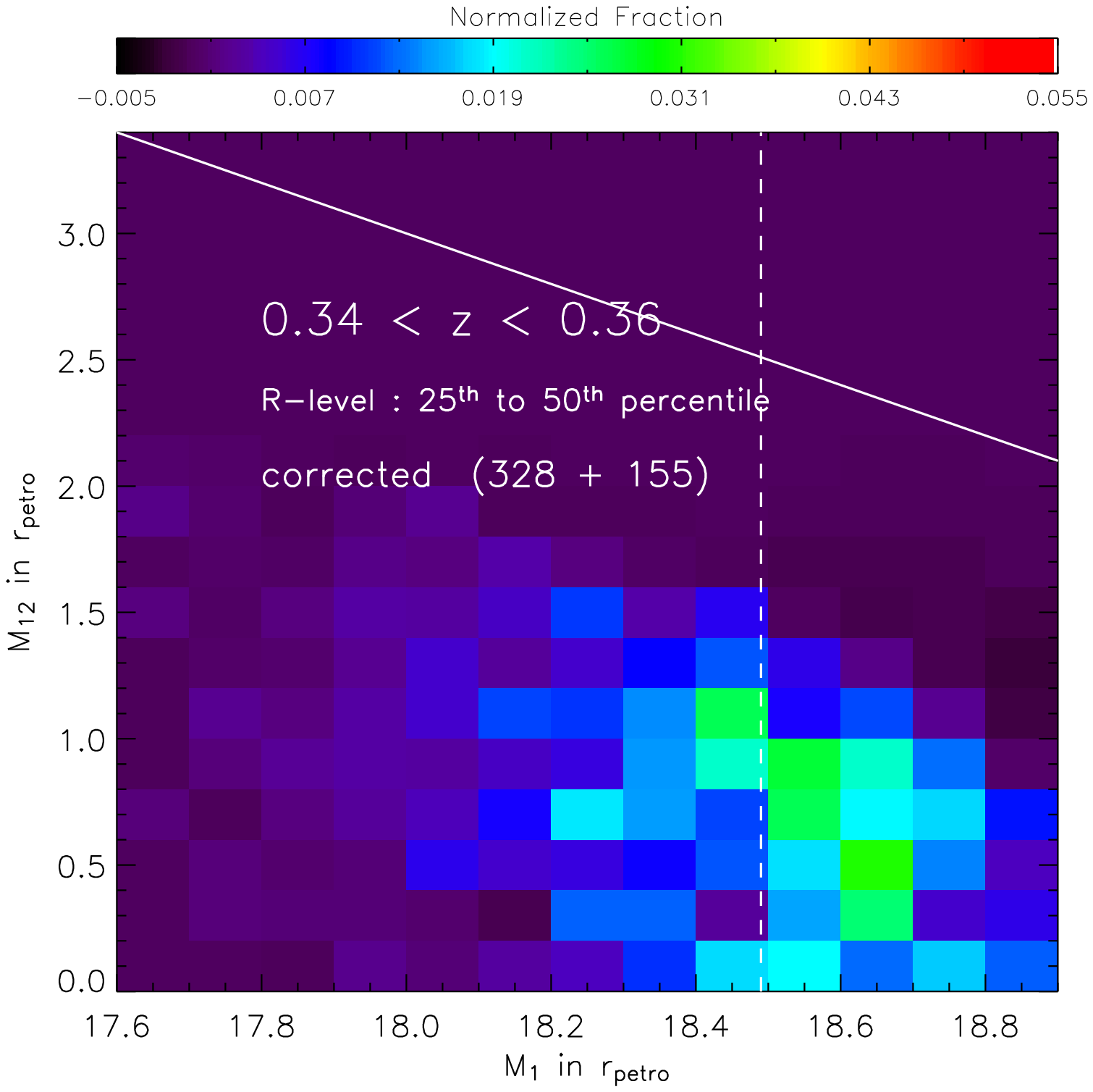}
\includegraphics[angle=0, width=0.48\textwidth]{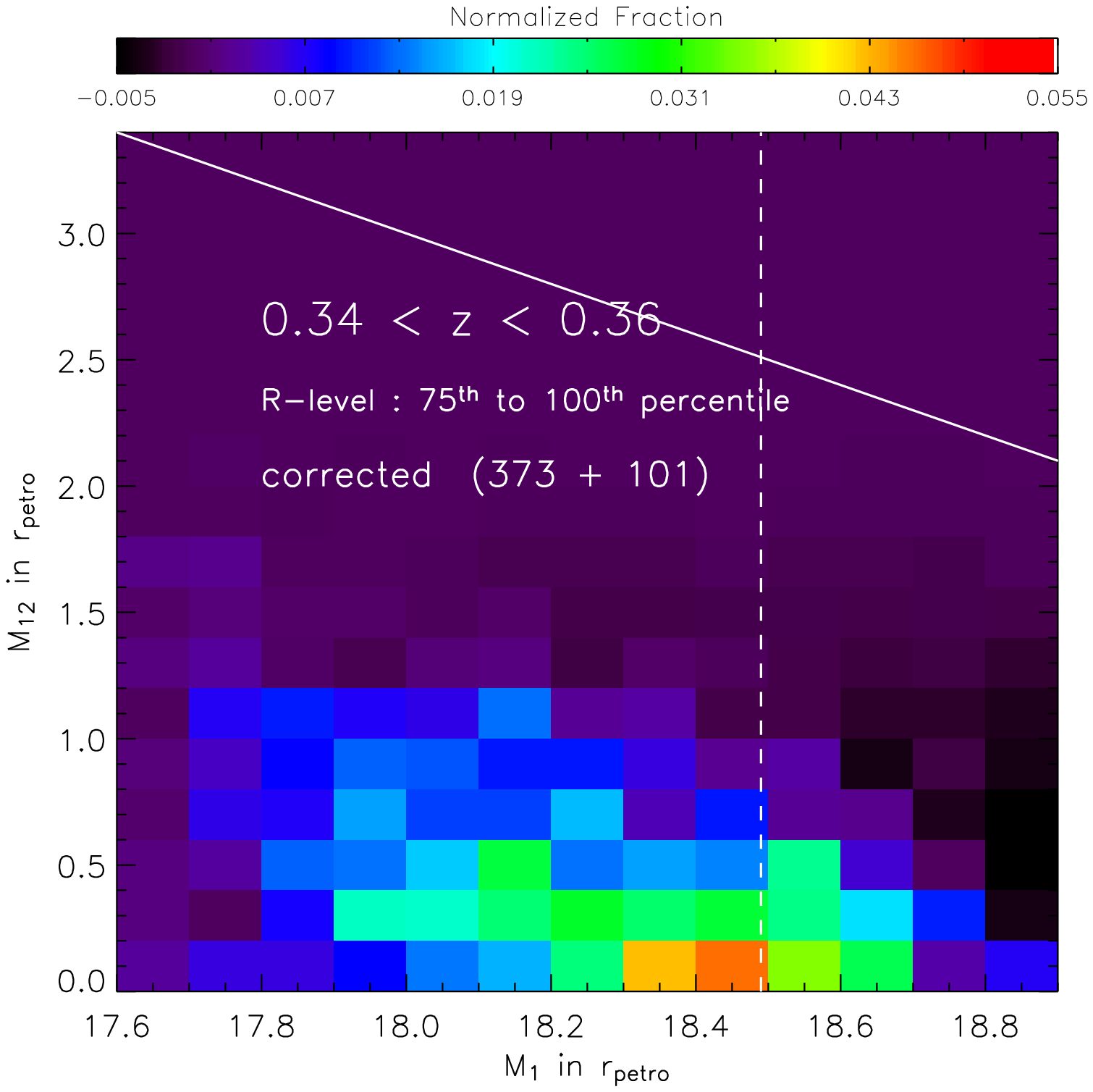}
\includegraphics[angle=0, width=0.48\textwidth]{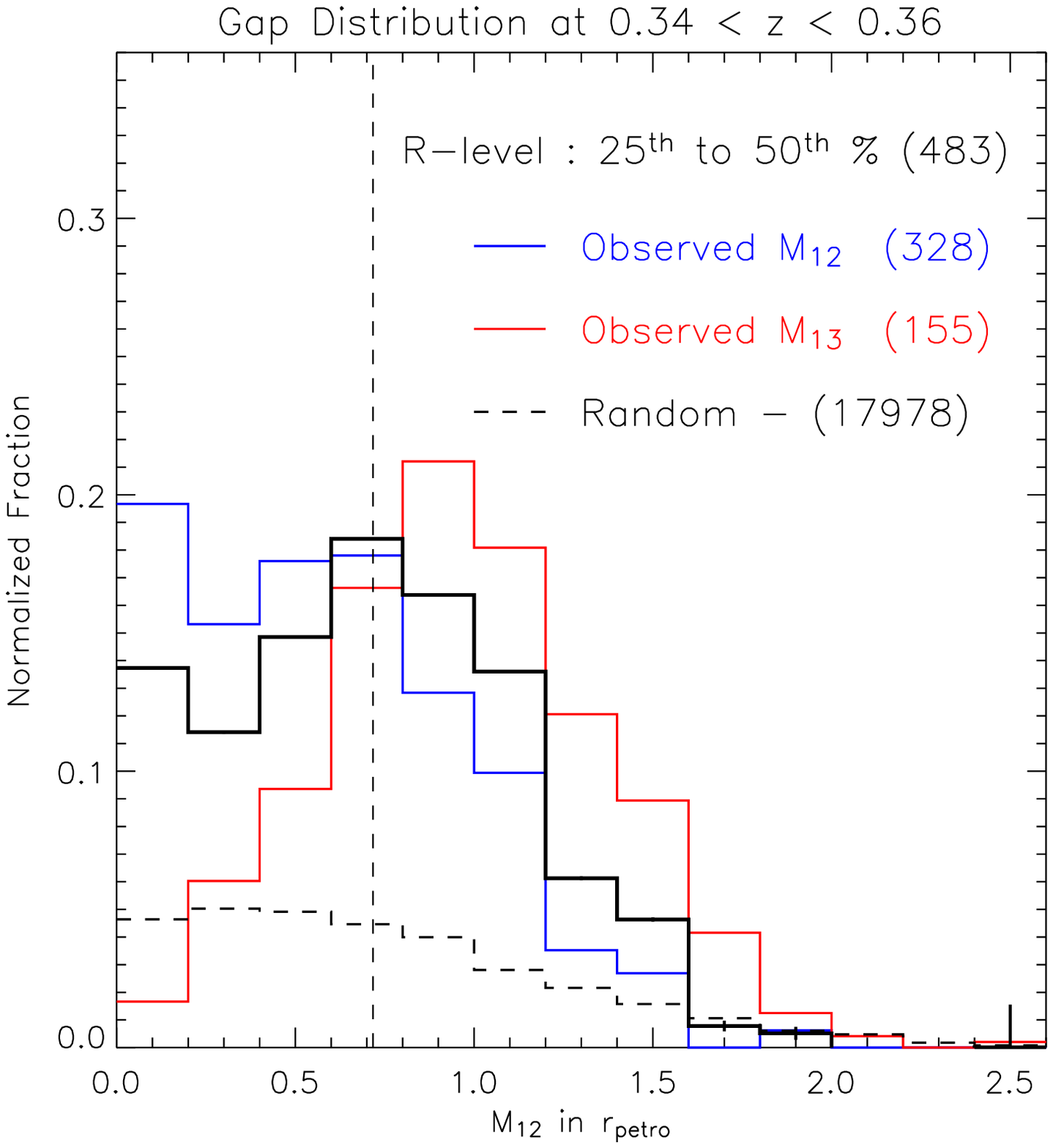}
\includegraphics[angle=0, width=0.48\textwidth]{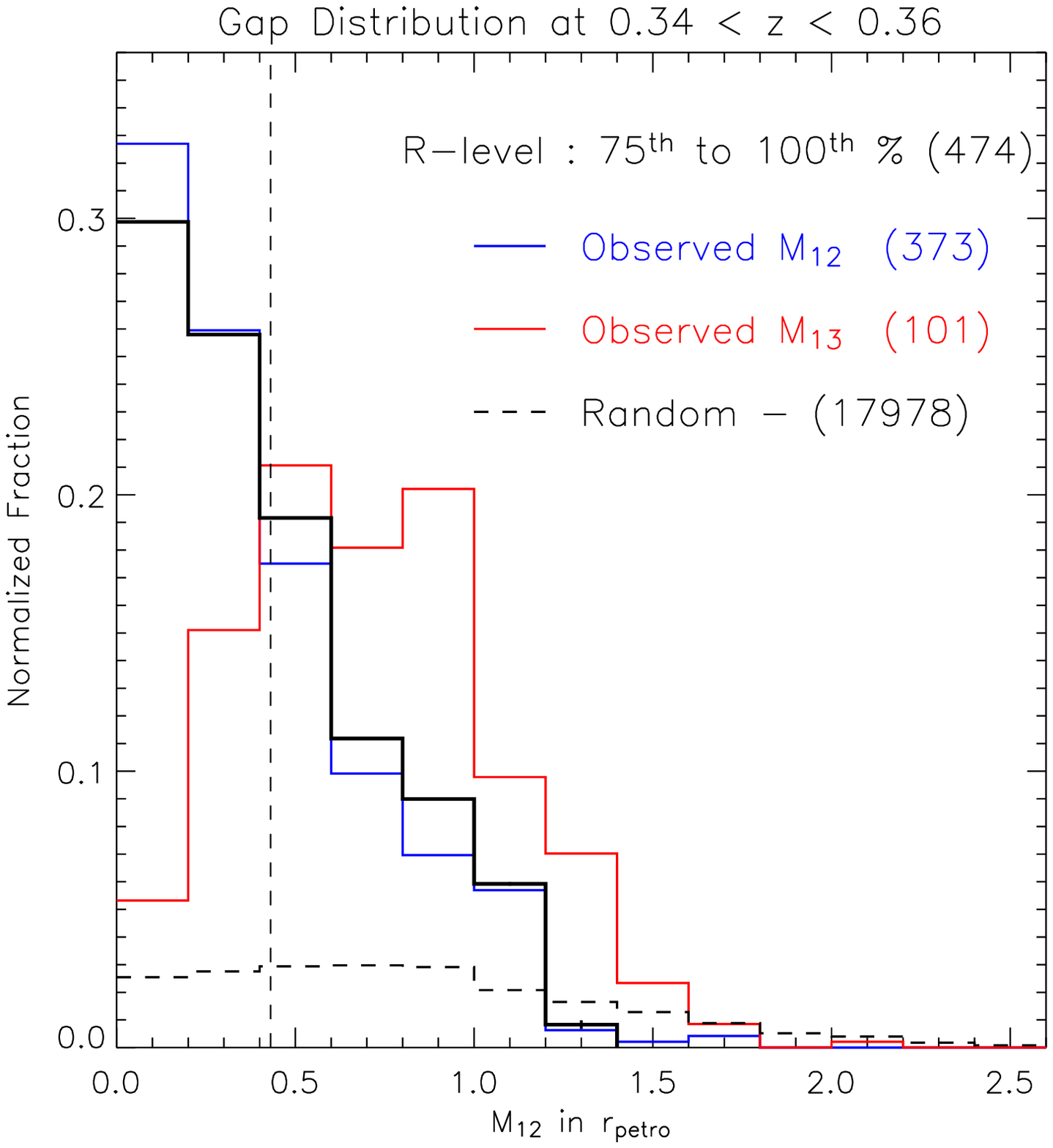}
\caption[]{As in Fig. \ref{fig:m12_hist_z0.15}, for redshifts $0.34 < z < 0.36$.
\label{fig:m12_hist_z0.35}}
\end{figure*}

This systematic smaller gap at higher redshift is also seen when the
sample is split into group-like ($25^{th}-50^{th}$) and cluster-like
(upper quartile) LRG fields, as shown in the top panels of Figures
\ref{fig:m12_hist_z0.15} and \ref{fig:m12_hist_z0.35}. The bottom
panels of each figure show histograms of the marginal gap
distribution, $\phi(M_{12})$, explicitly showing the noise level, 
$\phi^{\rm back}(M_{12})$ (dashed) determined from 10 random positions
for each LRG, the observed $\phi^{\rm obs}(M_{12})$ (blue) and
$\phi^{\rm obs}(M_{13})$ (red), and the final distribution
$\hat{\phi}(M_{12})$ (solid) used for our analyses as estimated from
equation (\ref{eq:gap_est}). The vertical dashed line is the mean
$\langle M_{12}\rangle$ for each panel, estimated from the first moment
of $\hat{\phi}(M_{12})$. It shows a slight decrease with redshift for
both clusters and groups.

The \cite{Bau70} effect, whereby clusters with brighter BCGs
have larger gaps, is seen in both rich and poor systems at all
redshifts. One measure of the strength of this effect is the tilt of
the number density ellipsoid of the joint distribution,
$\phi(M_1,M_{12})$.  From Figures \ref{fig:m12_hist_z0.15} 
and \ref{fig:m12_hist_z0.35}, it appears that this effect is stronger in
poorer systems, but changes little with redshift.

\begin{figure*}
\noindent
\includegraphics[angle=0, width=0.482\textwidth]{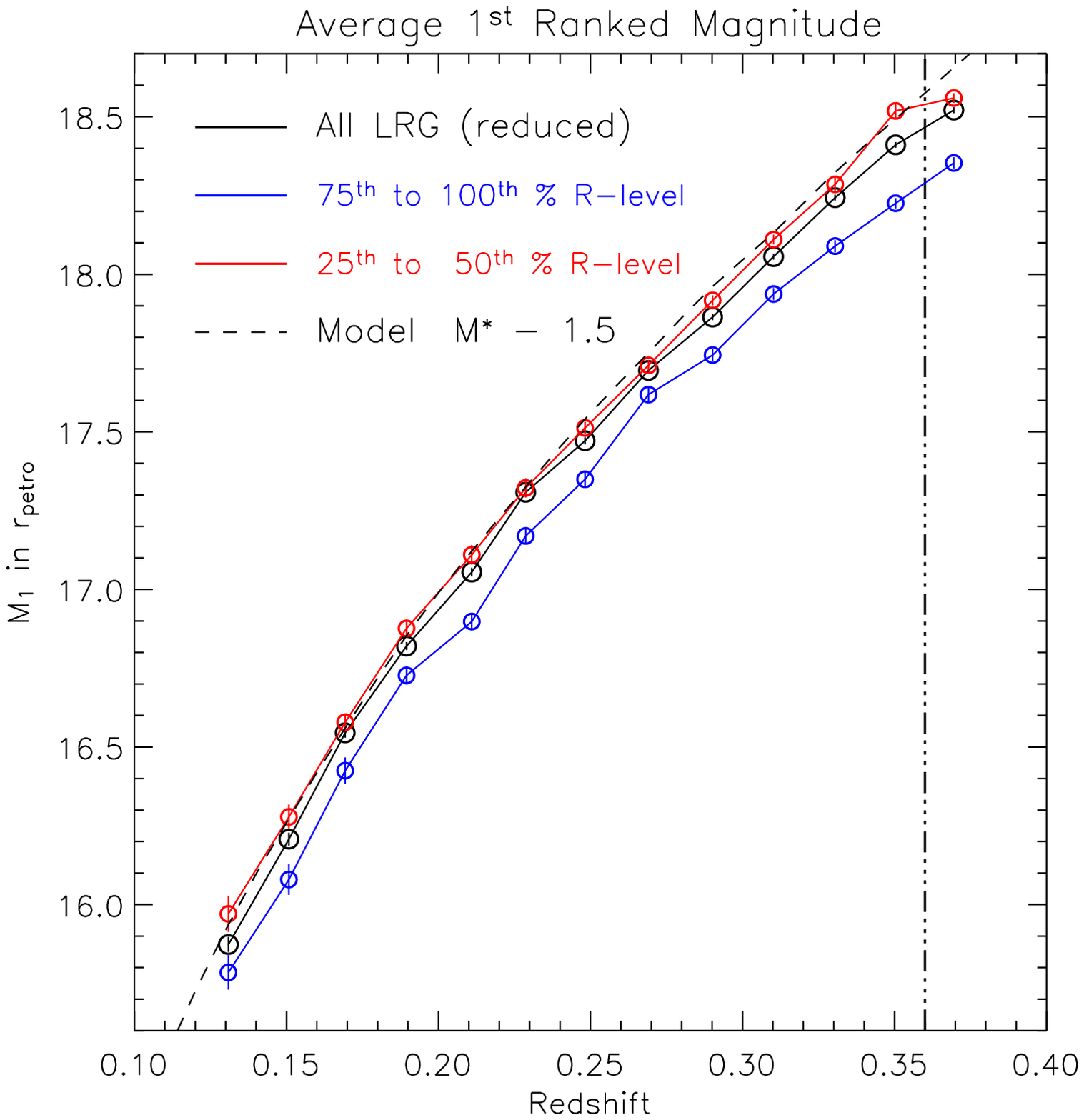}
\includegraphics[angle=0, width=0.482\textwidth]{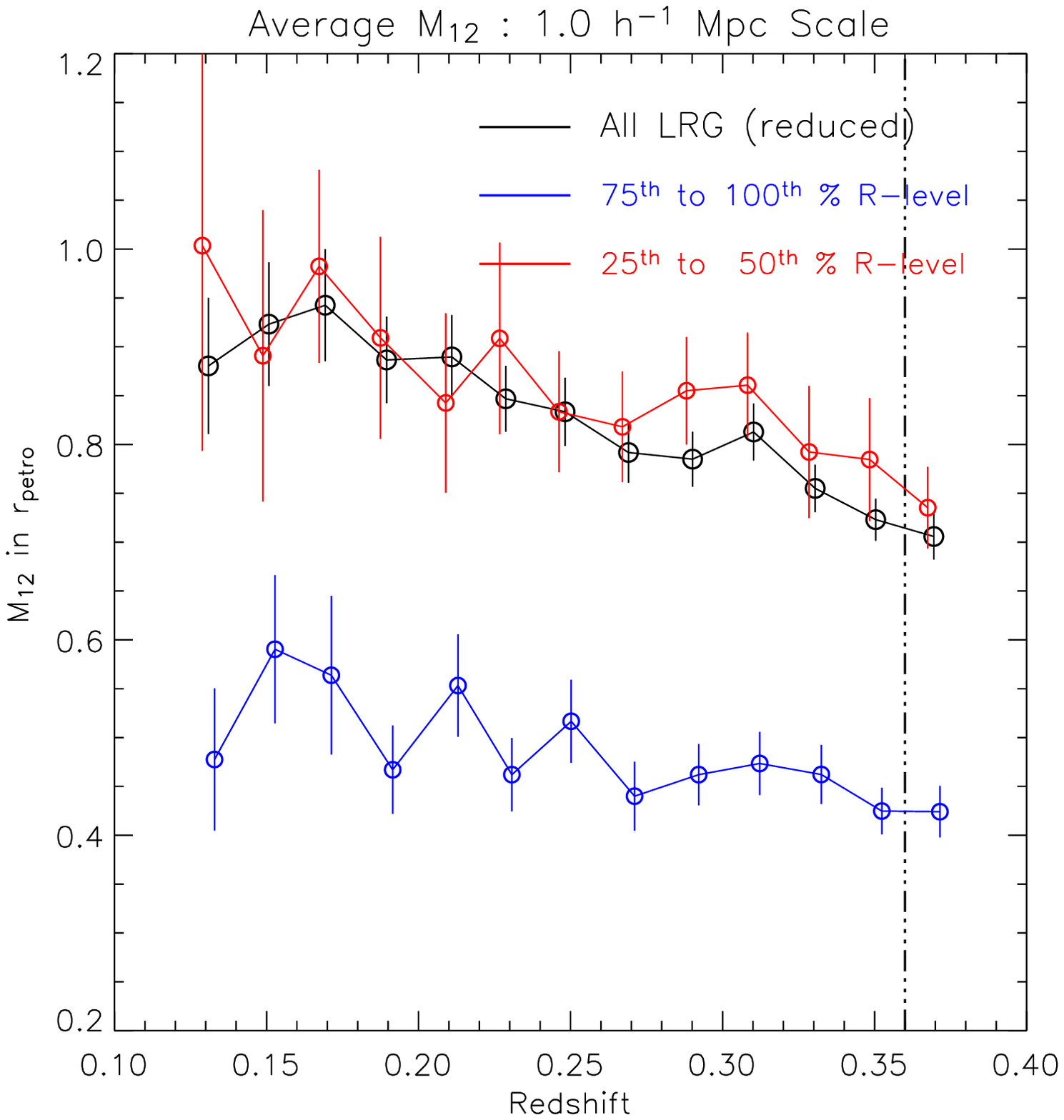}
\includegraphics[angle=0, width=0.482\textwidth]{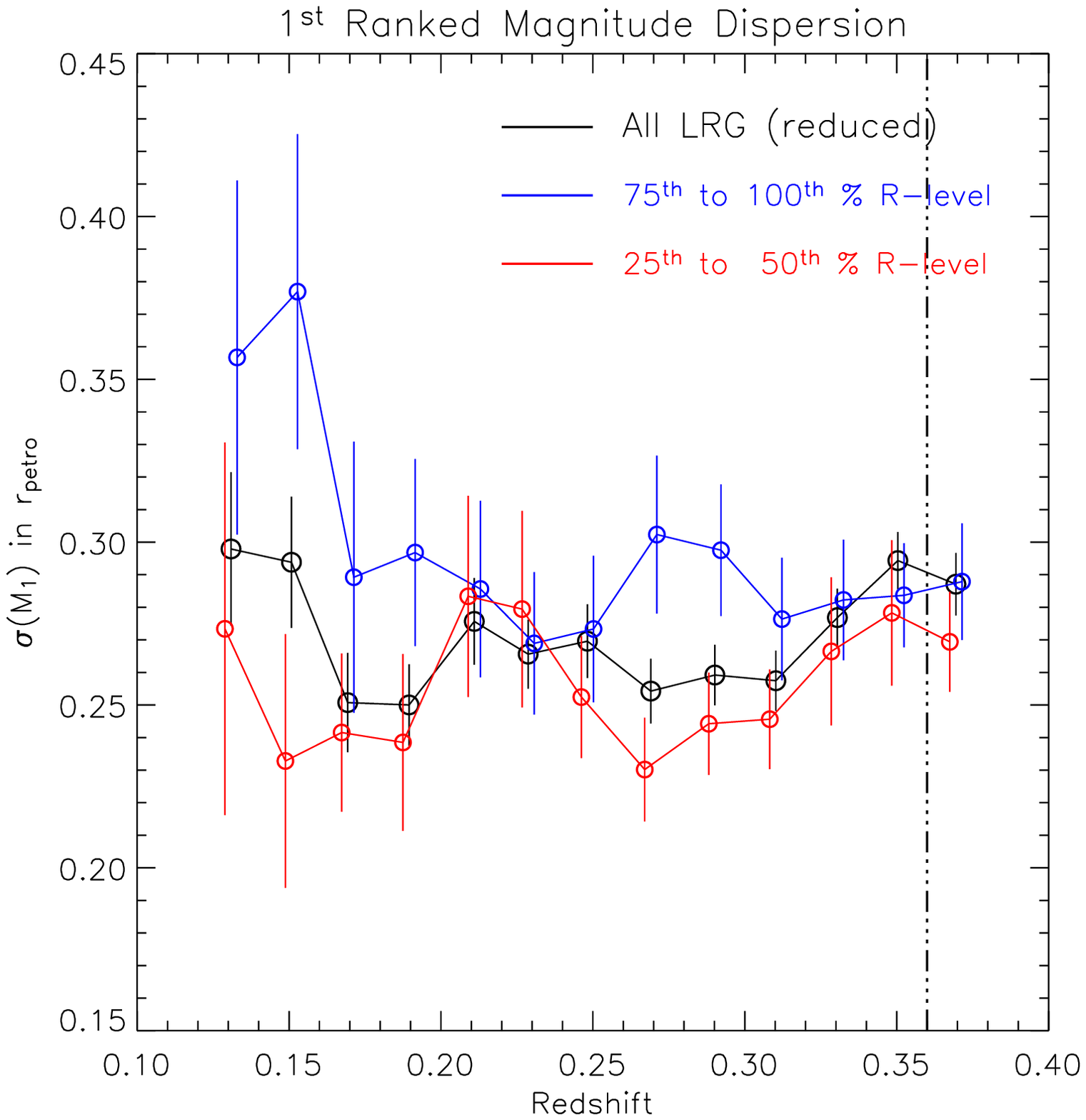}
\includegraphics[angle=0, width=0.482\textwidth]{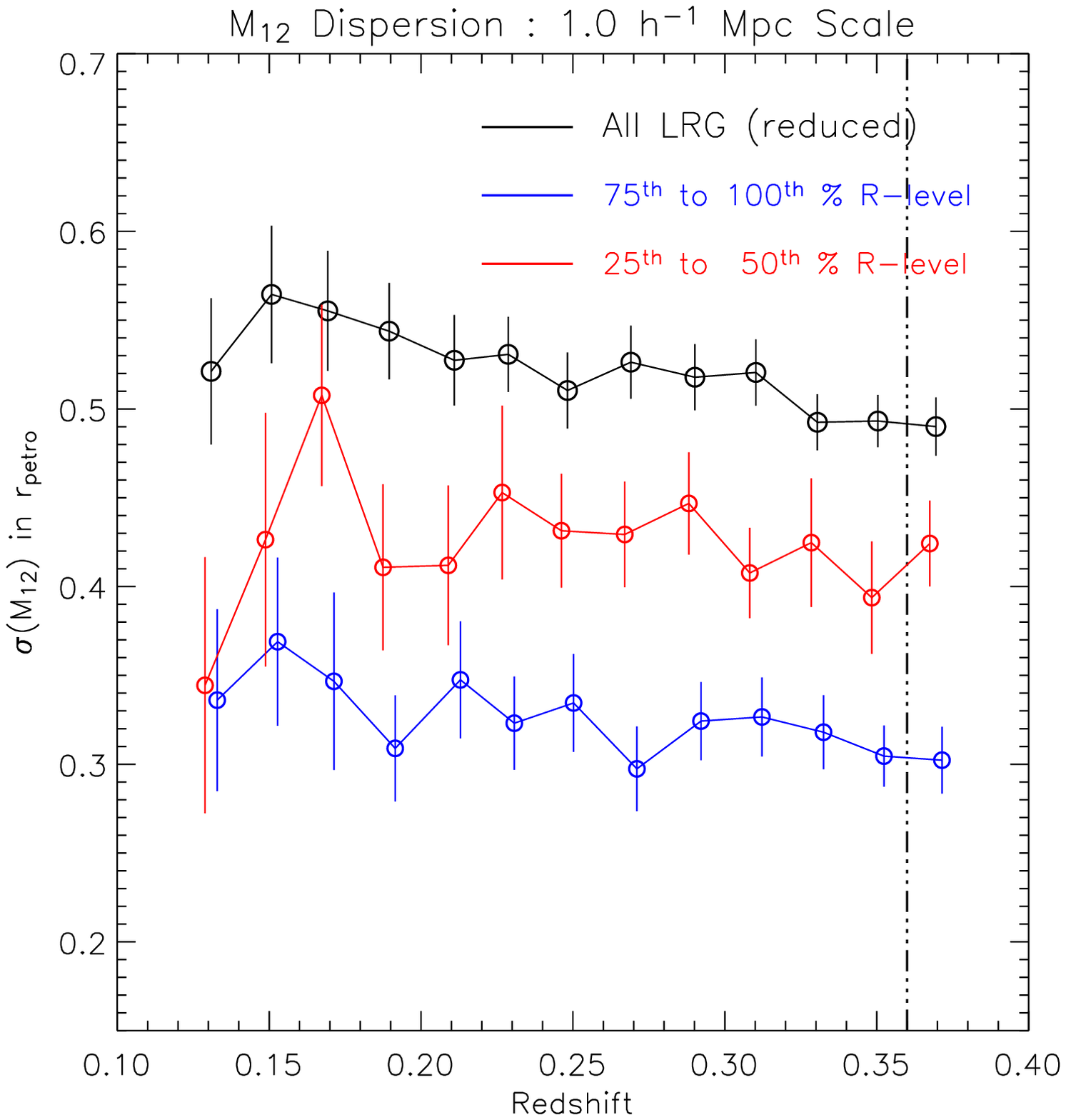}
\caption[]{(Left Panels) Average $1^{\rm st}$ ranked $r_{\rm petro}$ 
magnitude (top) and its corresponding 1 $\sigma$ dispersion (bottom) as a function of redshift.  
(Right Panels) $\langle M_{12} \rangle$ and its corresponding 1 $\sigma$ dispersion 
as a function of redshift. The blue (red) is for fields with richness in the upper 
$75^{\rm th}$($25^{\rm th}-50^{\rm th}$) percentile. Errors bars are $1 \sigma$
from bootstrap. The dot-dashed line is at $z = 0.36$, the redshift beyond which the sample is no 
longer volume-limited.\label{fig:mag_z}}
\end{figure*}
\begin{figure*}
\includegraphics[angle=0, width=0.482\textwidth]{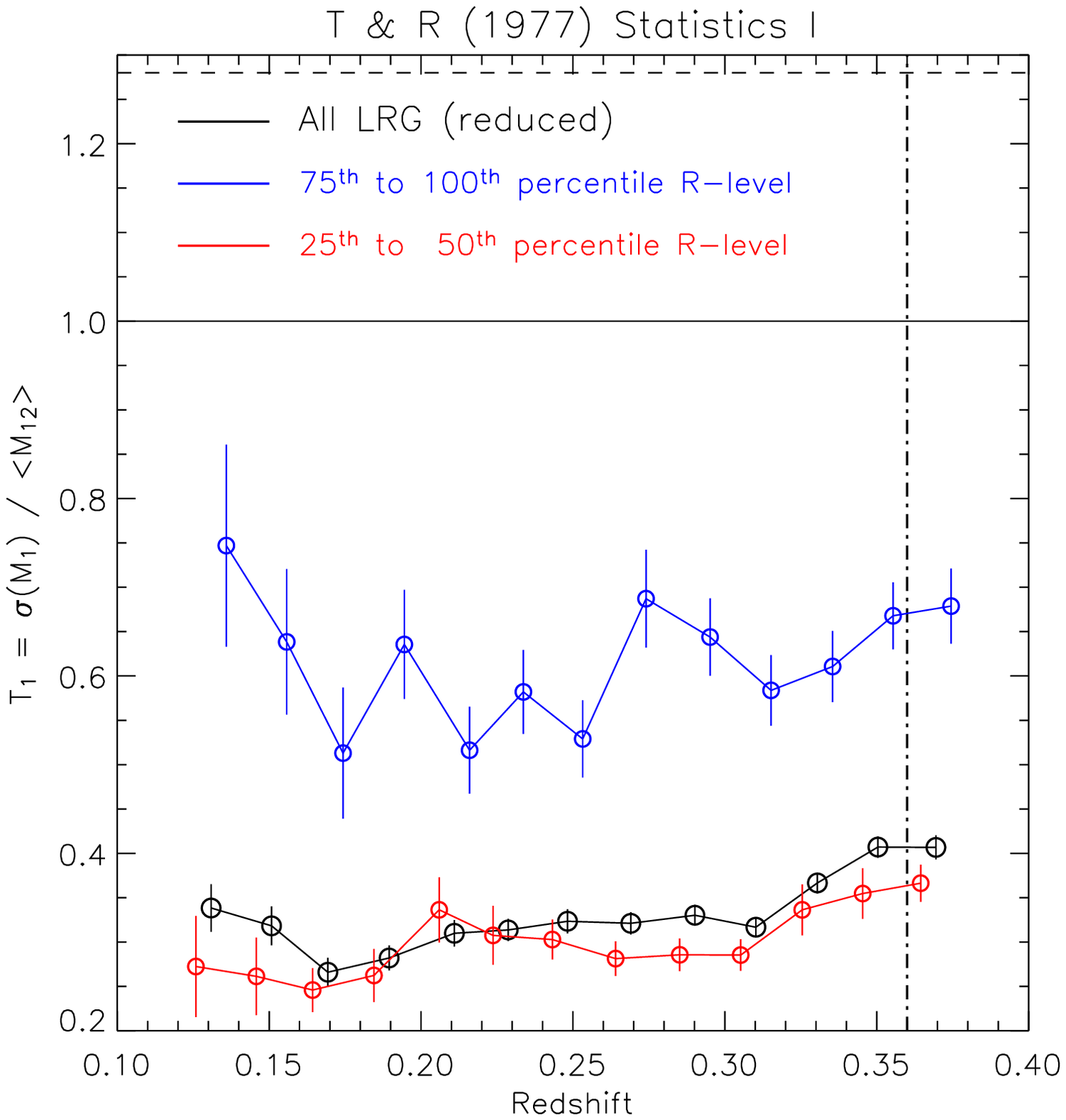}
\includegraphics[angle=0, width=0.482\textwidth]{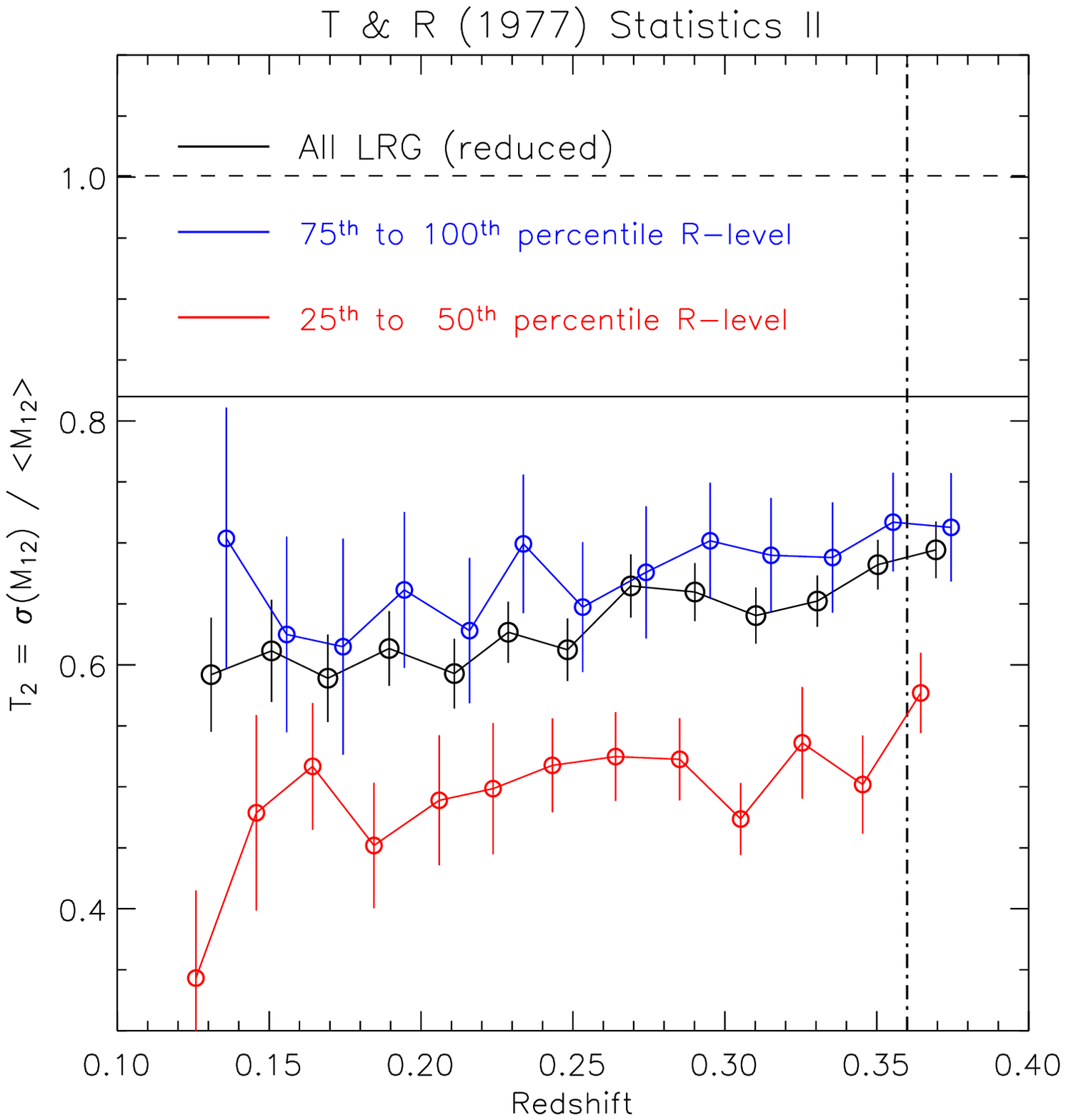}
\caption[]{The \cite{Tre77} statistics $T_1$ and $T_2$ as a function of redshift
computed in redshift slices $\Delta z = 0.02$. The solid horizontal line 
gives bounds below which brightest members would cease to be the statistical
extreme of the luminosity function of lesser members in the cluster. The top
dashed line is the prediction from extreme-value statistics. 
The blue (red) curve is for fields with richness in the upper 
$75^{\rm th}$($25^{\rm th}-50^{\rm th}$) percentile. Errors bars are 
$1 \sigma$ from bootstrap. The dot-dashed line is at $z = 0.36$, 
the redshift beyond which the sample is no longer volume-limited.
\label{fig:t1_t2_z}}
\end{figure*}
LRG fields with group-like environments have a larger gap than do 
cluster-like environments at all redshifts. This is seen in both the
joint distribution, $\phi(M_1,M_{12})$, and marginal distribution, 
$\phi(M_{12})$ (Figures \ref{fig:m12_hist_z0.15} and \ref{fig:m12_hist_z0.35}). 
This must in part be statistical: 
Poisson sampling from a common luminosity function naturally gives a 
smaller gap for a larger total number of galaxies. 
We will see below from the Tremaine-Richstone statistics that this is not the
whole story; Figure \ref{fig:t1_t2_z} shows that the bounds of  
equations (\ref{eq:t1}) and (\ref{eq:t2}) are violated. 
If the universal galaxy luminosity function estimated over 
large scale \citep[cf. ][]{Bla03} is correct, i.e. the bright end 
cuts off exponentially like a \citeauthor{Sch76} function, then the Poisson 
assumption must not hold. 

Figure \ref{fig:mag_z} summarizes the evolution of the mean and 
dispersion of $M_1$ and $M_{12}$ 
from the various sub-samples. The top right panel
shows that poorer systems have first-rank members that are brighter by
0.1--0.2 mag in the mean, while the top left panel shows that these
same systems have a larger mean $\langle M_{12}\rangle$ at all redshifts. 
Our sample is volume-limited independent of richness, and the second and third ranked 
galaxies are also complete for the range of $M_{12}$ and $M_{13}$ we are probing,
given the depth of the SDSS photometric data.  The dependence of the
gap size on richness is not an artifact of 
our richness estimates, as the mean (and median) number of $M^*$ and brighter 
galaxies is roughly constant with redshift, after background
subtraction (cf. \citealt{Loh05b}). 

Our results for the \citeauthor{Tre77} statistics $T_1$ and $T_2$ are
presented in Figure \ref{fig:t1_t2_z}. The estimated $\hat{T}_1$ and
$\hat{T}_2$ are well below the lower bounds for the
statistical BCG hypothesis (indicated by the solid horizontal
line) set by equation (\ref{eq:t1}) and (\ref{eq:t2}) for both poor
and rich systems at all redshifts.  They are even further away from
the statistical i.i.d. prediction (upper dashed line) from
extreme-value statistics of \cite{Bha85}. If we assume that the bright
end of the galaxy luminosity function in clusters cuts off
exponentially like the universal galaxy luminosity function,
then the statistical hypothesis for BCG is rejected with high
confidence, even if we allow the steepness of this cut-off to vary
from cluster to cluster.  Of course, this result has meaning only if
our initial assumption about the galaxy luminosity function embodied
in \citeauthor{Sco57}'s model -- of independent luminosity sampling --
is valid. The assumption that the luminosities of galaxies are
independent must break down at some level, since we know that galaxies
influence with each other via tidal interactions and mergers through 
gravity to very large distances. On the scale of $1.0\,\hmpc$, 
the independence assumption is unlikely to be true especially 
for galaxies as massive as LRGs ($\sim 5-8\times 10^{12} M_{\odot}$). 
Hence the fact that $T_1$ and $T_2$ fall well below the boundaries of 
equations (\ref{eq:t1}) and (\ref{eq:t2}) does not by itself imply
that BCGs follow 
an assembly path different from that of other cluster galaxies.
It might merely signal a breakdown in our assumption about the galaxy 
luminosity function. Indeed, bright elliptical galaxies, including 
BCGs \citep{Oeg91,Pos95,Gra98} form a low dimensional sequence (the Fundamental
Plane relation, \citealt{Dre87}; \citealt{Djo87}), implying some kind
of universal  assembly history. 

$T_1$ and $T_2$ are the inverse of the mean luminosity ratio of the two
brightest galaxies, measured in units of the two characteristic
luminosity spreads, $\sigma(M_1)$ and $\sigma(M_{12})$.  When
considered as a function
of redshift, they quantify the \emph{normalized} relative
brightening of the first-ranked galaxy, as compared with lesser 
members of the same cluster. These normalizations are with respect to the 
dispersions of $M_1$ and $M_{12}$.  
$\hat{T}_1(z)$ and $\hat{T}_2(z)$ are consistent with being flat
as a function of redshift, suggesting that there is no 
major process that influences the creation of the observed luminosity 
gap over the redshift range we probe.  This can be compared with the raw gap
function, $\langle M_{12}\rangle(z)$, of the top left panel in Figure
\ref{fig:mag_z}, where a decrease with redshift is seen, suggesting
mild brightening of the first-ranked galaxy compared to the
second-ranked galaxy with time. Nonetheless, the corresponding spread
in magnitude, notably $\sigma(M_{12})$, also decreases with
redshift, in such a way that $\hat T_2$ remains flat.  

\section{Discussion}
\label{sec:discussion}
\subsection{Interpretive Caveats}
Our richness measure says little about the gravitational potential of
the LRG fields, since we do not measure the mass overdensity directly.
Our proxy for overdensity, the $R$-level yardstick, is tuned to
robustly characterize the environment of rich systems with high
signal-to-noise ratio. Our groups (i.e., the $25^{\rm th} - 50^{\rm th}$ percentile sample)
by design must host a luminous ($\la M^\ast - 1$ or brighter) LRG, hence it is a highly biased
sub-class of galaxy groups. The comoving number density of our groups is a mere $2 \times 10^{-5} \ihmpcC$, 
an order of magnitude below the expected group density from the analytical 
\cite{PS74} estimates.
Hence, we do not sample a substantial fraction of group-like
environments in the universe. 
Even if the mass--to--$R$-level scaling relation is tight and 
unbiased \citep{Loh05c}, our richness measurements 
can give biased mean mass estimates in the present of moderate intrinsic 
scatter and measurement uncertainties. Further, it could be that our groups 
are equally massive, with a huge dwarf galaxy population compared to their richer
counterparts, as observed locally in a subset of X-ray bright Hickson compact groups \citep{Hun98}.

These groups are higher redshift analogs of the famous
\cite{Mor75} and \cite{Alb77} groups that host luminous (non-c) D
galaxies. However, we  are only sampling a subset of special
groups, which may resolve the apparent contradiction between our results and
those of \cite{Gel83} -- they found that the brightest members of
groups are less dominant than those of clusters and are consistent
with them simply representing the luminous tail of the luminosity
function. It would be of interest to know what fraction 
of \emph{all} groups (observationally defined) host an LRG. We will show in
a subsequent paper \citep{Loh05c} that $\sim 95 \%$ of the most massive 
clusters selected by both the optical matched-filter techniques 
\citep{Kim02a} and X-rays have LRGs in them, but the fraction for groups 
is expected to drop substantially. Note that the number density of groups
from optical redshift surveys have a higher number density than LRGs, in part
because many of these groups are not actually virialized and display no
extended X-ray emission. Further, there are groups with extended X-ray emission
that have bright elliptical members that are just shy of the $M^* - 1$ luminosity
threshold of the LRG \citep{Nip03}. 

Another caveat of our analysis is the $1.0\,\hmpc$ scale used to
search for second and third ranked galaxies. If we choose a smaller
scale, say $0.25 \hmpc$, we find a larger mean gap for both
richer and poorer system, by $\sim 0.5$ {\tt mag}. However, the
difference in the gap between rich and poor systems is smaller. For
example, at the current $1.0 \hmpc$ scale, the average difference of
the estimated $\langle M_{12} \rangle$ between the rich and poor
systems is $0.4$ {\tt mag} (see the top right panel of Figure
\ref{fig:mag_z}), while at $0.25 \hmpc$ it is $\sim 0.15$ {\tt mag}.
This difference is again in part Poissonian, but is also influenced
by how bright galaxies populate a cluster. The smaller scale analysis
of $0.25 \hmpc$ would, for example, split the Coma cluster into two
LRG fields, each with a large luminosity gap, while the
present $1.0 \hmpc$ analysis would treat Coma as a single system with a
small gap. We could try to infer a dynamically motivated scale
like a virial radius or $r_{\rm 200}$ -- the radius at which the galaxy
density drops below 200 times the mean density -- but these measurements
are noisy and ill defined for poor systems.

\subsection{Comparison with Numerical Studies}
We compare our findings with the numerical experiment of \cite{Dub98},
who simulated the formation of groups and poor clusters of galaxies
in a cosmological setting (albeit based on the now out-of-favor
$\Omega_m=1, \sigma_8 = 0.7$ CDM model).   Our $25^{\rm th} - 50^{\rm
  th}$ percentile 
$R$-level sample matches his systems. His simulation specifically
addresses the growth of the brightest galaxy and its creation in the
context of cluster collapse in a hierarchical structure formation
paradigm. His experiment begins at $z = 2$, and by $z \sim 0.8$ the
major galactic unit that is later identified as the BCG is already in
place. This BCG is a result of mergers of the brightest few galaxies
of roughly equal mass during the triaxial collapse of the cluster.
By $z \sim 0.4$, all major mergers between the BCG and other
galactic units are complete. The large mass contrast between the BCG and the
second-ranked galaxy is now in place. This would suggest that $M_{12}$ will
shrink at $z > 0.4$, unfortunately beyond the redshift which our
sample probes. Qualitatively, our results from
the group-like LRG fields are in agreement with his numerical
experiment. The large and unchanging gap seen in these poorer
systems points to an earlier origin of the growth of the
dominance of the BCG, of which the picture painted by his
experiment is one plausible description.   However, cluster infall and
growth is a strong function of $\Omega_m$ and $\sigma_8$ (in
particular, models with smaller $\Omega_m$ predict infall ending at
much higher redshift).  Thus there is an urgent need for
simulations of BCG growth using the concordance cosmology. 

\subsection{Are First-Ranked LRGs drawn from the Universal Luminosity Function?}
BCG show a large luminosity gap; moreover, the dominance of the BCG is
observed to be independent of redshift.  Thus suggests that whatever
mechanism created this dominance must
have taken place at $z > 0.4$. Scenarios in which recent cannibalistic
activity fueled the BCG growth \citep[e.g.][]{Hau78} are
unlikely.  Poorer systems have more dominant BCGs, suggesting that
the process that makes these galaxies unusually luminous
is not related to the presence of BCGs in crowded rich cluster cores. 
Using 211 clusters of galaxies selected from SDSS
photometry, \cite{Kim02b} found that BCG with $M_{12} > 0.5$ 
{\tt mag} showed strong alignment with their host clusters, while
those with weaker dominance showed no tendency to align. 
(See also \cite{Ful99} for similar results for MKW/AWM groups.)
Taken together, these observations support the scenario of 
\cite{Dub98} -- essentially a
generic feature of the hierarchical structure formation paradigm with
Gaussian initial conditions -- whereby major mergers during galactic
infall along filaments create the BCG at moderate ($z \sim
0.5\,-\,1.0$) redshifts.  Statistical sampling may explain why poorer
systems (which host an LRG) have more dominant brightest members.
Alternatively, these systems are preferentially located within older
structures, and are dynamically more mature. The BCGs at the confluence
of these structures are further along the evolution sequence, making them 
more dominant. Nevertheless, it suggests that the properties of BCGs are 
intimately related to the assembly history of the local matter density. 
These findings, however, run contrary to the standard candle nature of 
the BCG metric luminosities \citep{San72,Gun75,Pos95}, and argue against a 
BCG special population independent of local conditions. Properties of BCG
clearly depend on environment and yet they are standard candles. We do not
have a full understanding of this dichotomy.

\section{Conclusions}\label{sec:gap_conclude}
We have shown that the luminosity gap, $M_{12}$ between the
first-ranked galaxy and the second-ranked in fields containing at
least one LRG is large, and this is inconsistent with these brightest
members being the statistical extreme of the local galaxy luminosity
function within \cite{Sco57}'s model. This dominance of the brightest
members changes little over the redshift range of $0.12 < z < 0.38$,
and is more prominent in the poorer group-like LRG fields than the
richer cluster-like fields. Furthermore, the brightest member in cluster-like
LRG fields are systematically brighter (0.1--0.2 {\tt mag}) in the mean
than in group LRG fields. When we combine these findings with the
conclusions of \cite{Kim02b} and \cite{Ful99} that dominant brightest members are 
preferentially aligned with their host clusters, our results support
the assembly of BCGs during cluster collapse within the framework of
hierarchical structure formation, and put stringent constraints on the
growth of BCGs through recent accretion of lesser members. 
In future papers, we will present a statistical analysis of the
optical environments of this volume-limited sample of LRG \citep{Loh05b}, 
and their relationship with intermediate redshift clusters selected by 
X-ray emission \citep{Loh05c}.

\section*{Acknowledgments}
We thank Rita Kim, Marc Postman, Tod Lauer, Jim Gunn, Scott Tremaine, 
Michael Blanton, David Hogg, Don Schneider, Tim Mckay, Richard Cool and 
Daniel Eisenstein for suggestions. 

Funding for the creation and distribution of the SDSS Archive has been provided 
by the Alfred P. Sloan Foundation, the Participating Institutions, the National 
Aeronautics and Space Administration, the National Science Foundation, the U.S. 
Department of Energy, the Japanese Monbukagakusho, and the Max Planck Society. 
The SDSS Web site is {\tt http://www.sdss.org/.}

The SDSS is managed by the Astrophysical Research Consortium (ARC) for the 
Participating Institutions. The Participating Institutions are The University of Chicago, 
Fermilab, the Institute for Advanced Study, the Japan Participation Group, 
The Johns Hopkins University, the Korean Scientist Group, Los Alamos National Laboratory, 
the Max-Planck-Institute for Astronomy (MPIA), the Max-Planck-Institute for Astrophysics (MPA), 
New Mexico State University, University of Pittsburgh, University of Portsmouth, 
Princeton University, the United States Naval Observatory, and the University of Washington.
YSL and MAS acknowledge the support of NSF grant AST-0307409. 

\bibliographystyle{mn2e}

\appendix
\section{A Consistent Population of LRG}\label{trimmed_LRG}
As discussed earlier in the main text and in \cite{Eis01}, the LRG target selection 
will only give a consistent population for $z > 0.23$. In particular, for $r < 17.5$, the 
selection admits many intrinsically fainter galaxies. Furthermore, bright galaxies are at the 
exponential tail of the luminosity function, and small deviations from the nominal LRG
luminosity threshold affect the derived number counts substantially.
Since the flux-limited MAIN galaxy survey is spectroscopically complete to 
$r \sim 17.8$ or $M^* - 1$ at $z \sim 0.25$, we spectroscopically trimmed the low 
redshift LRG to match the rest-frame color and luminosity of their higher redshift counterparts. 
We use the empirically derived color-redshift relation of $M^* - 1$ red sequence galaxies 
from Appendix \ref{appendix} to extract a co-evolving population of similar color and luminosity.  

\begin{figure*}
\includegraphics[width=0.481\textwidth,height=0.43\textwidth]
{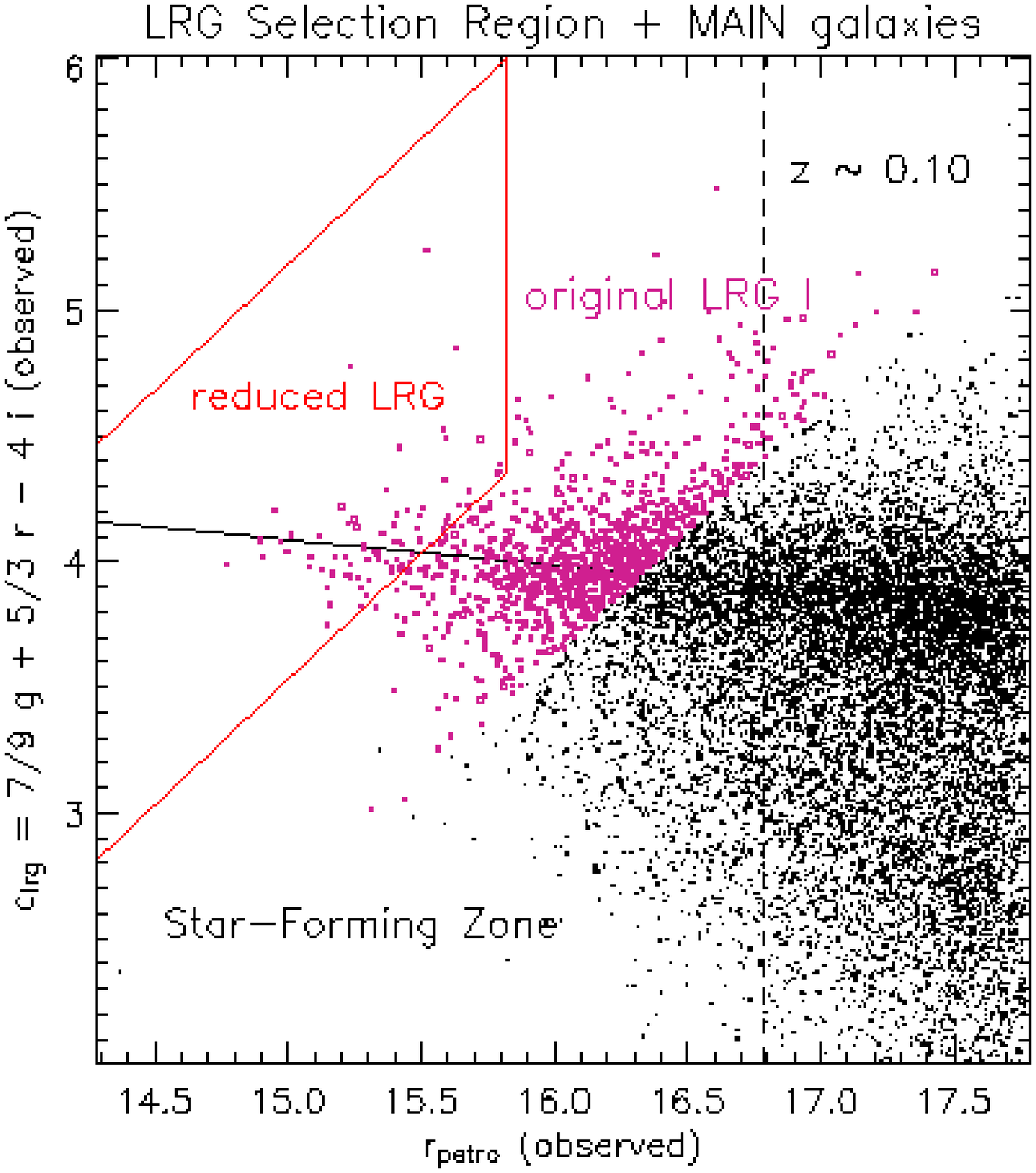}
\includegraphics[width=0.481\textwidth,height=0.43\textwidth]
{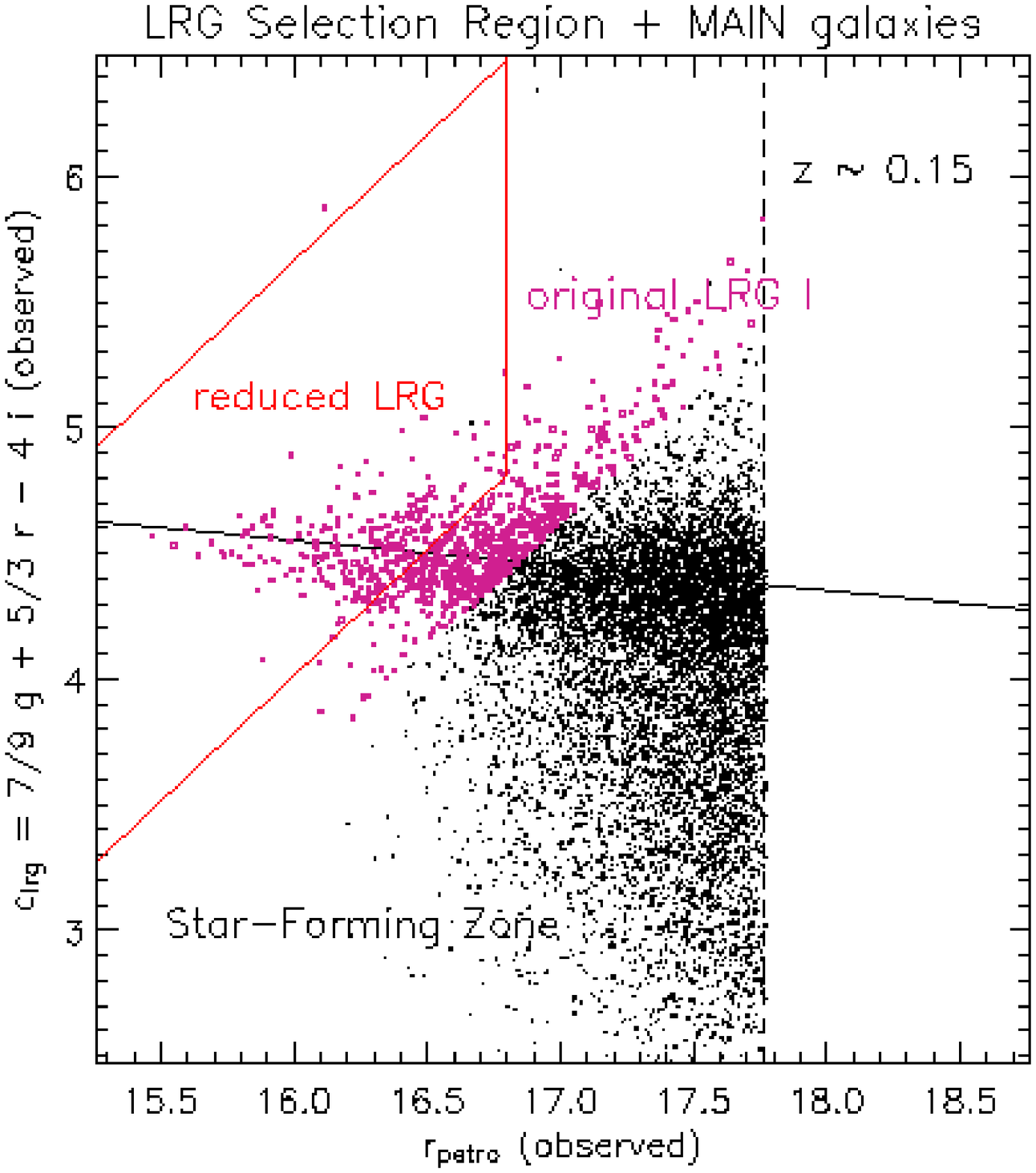}
\includegraphics[width=0.481\textwidth,height=0.43\textwidth]
{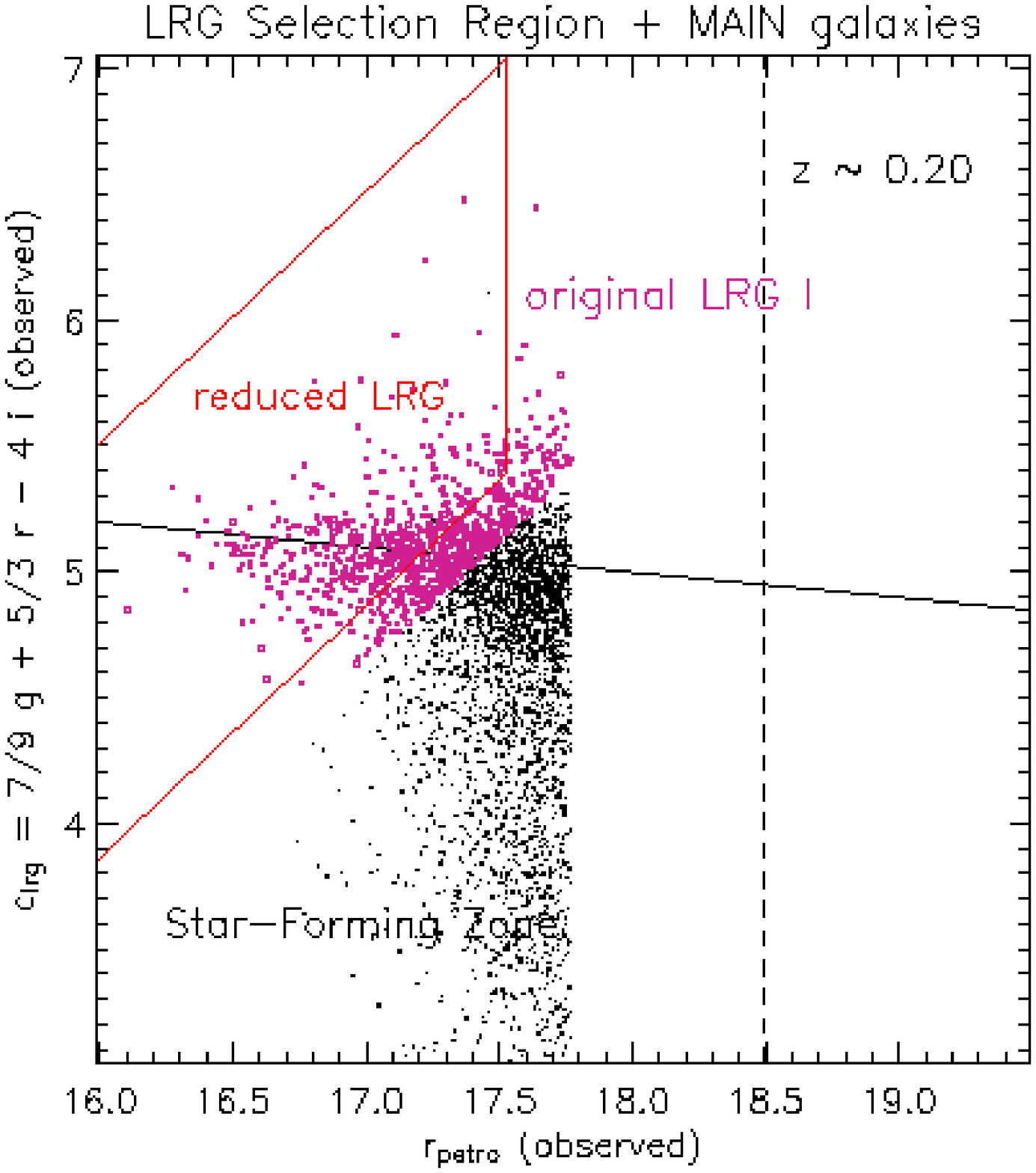}
\includegraphics[width=0.481\textwidth,height=0.43\textwidth]
{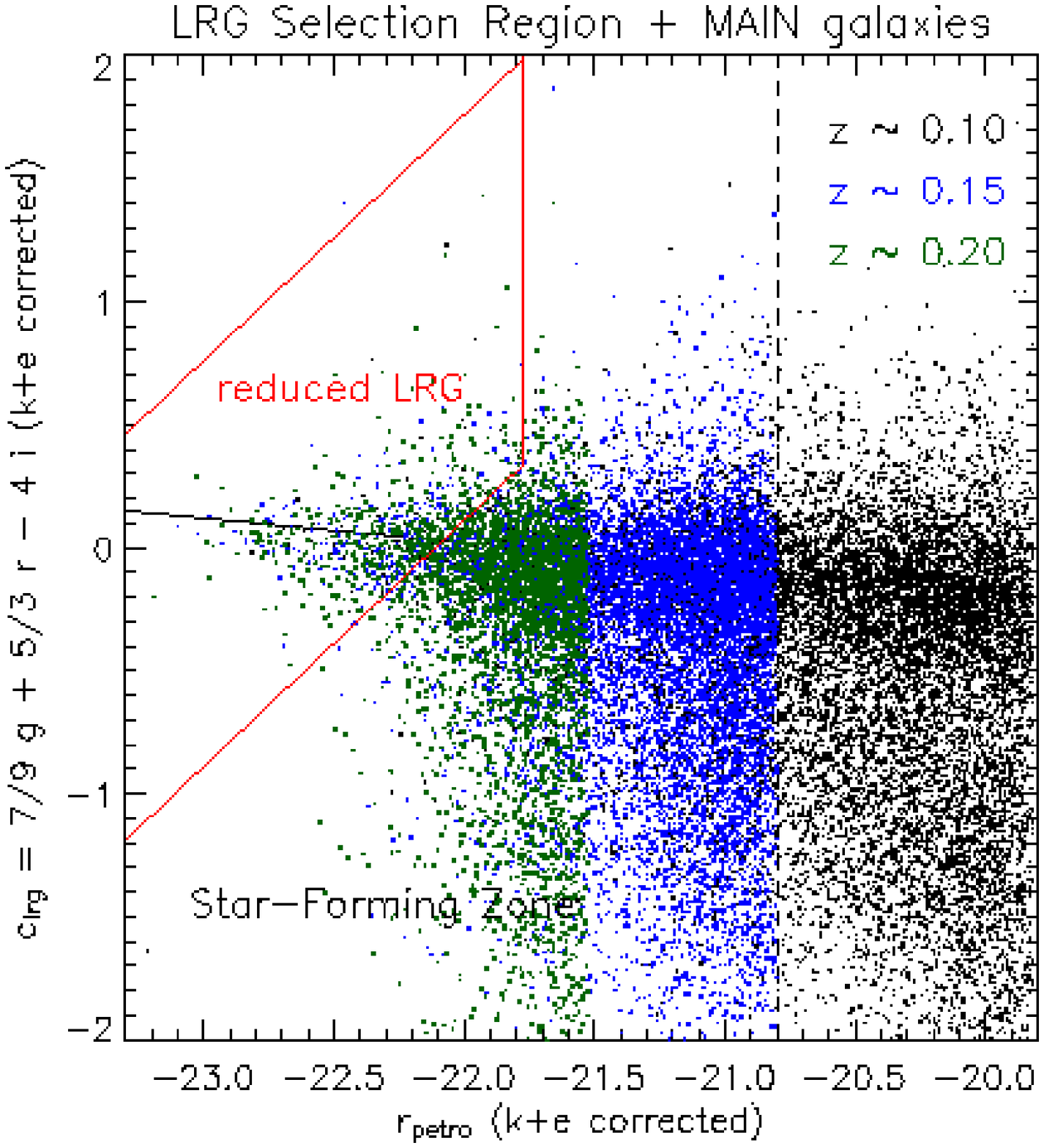}
\caption[]{\label{fig:main_lrg_region}
These panels show the observed color magnitude diagrams of LRG and MAIN sample galaxies
in three redshift ranges. The axes for the two top and the bottom left panels are in 
observed quantities. A model \emph{k+e} correction for an old stellar population 
sequence is used to shift the x-axes to the same luminosity range. $M^\ast$ at the 
respective redshifts of (0.1, 0.15, 0.2 ($\Delta z = 0.01$)) of the panels are aligned, 
and is indicated by the dashed vertical line. 
The location of the red sequence -- the sloping horizontal line -- 
for a fixed absolute magnitude is used to shift the y-axis to the same rest-frame color. 
The diagrams are made using spectroscopic galaxies brighter than $\rpet \sim 17.8$ 
from the MAIN sample. Hence, they are unbiased with respect to color. The violet points are 
these MAIN galaxies that also pass the LRG Cut I selection criteria. 
The red box is the rest-frame color-magnitude cut used here for the consistent trimmed LRG 
sample. The last panel (bottom right) plots all galaxies from the other three panels, 
now shown in rest-frame quantities.}
\end{figure*}

Figure \ref{fig:main_lrg_region} shows plots of the color-magnitude relation 
for galaxies from the SDSS MAIN galaxy survey.  The color used is 
$\clrg = 7/9 g + 5/3 r - 4 i$, which is approximately parallel to the color locus of 
LRGs; see Figure \ref{fig:gr_ri_locus}. The figure shows three narrow redshift slices 
($\Delta z = 0.01$) at $z = 0.10, 0.15$, and $0.20$. 
The red sequence is clearly visible. The axes for the 
first three panels (both top and bottom left) are labeled in \emph{observed} quantities, 
but shifted relative to the red sequence for direct comparison between the redshift subsamples. 
The fourth panel (bottom right) is a combined plot of the three, where the axis is now labeled 
in rest-frame quantities. We employ a single model {\it k+e} correction of a 
passively evolving early-type $M^*$ galaxy from PEGASE to realign the $\rpet$-axis for 
rest-frame comparison. Galaxies that pass the LRG target selection are indicated by violet
points. The steep oblique selection edge is the color-luminosity bound that imposed a
red bias to LRGs. Notice that the selection is more permissive at $z = 0.1$, but 
becomes gradually more stringent at $z \approx 0.2$\footnote{\normalsize{As 
an interesting aside, notice that even though the LRG target selection is permissive at 
$z \approx 0.10$, it still primarily selects galaxies on the red-sequence, as 
contamination from blue star-forming galaxies is minimal. Thus one could in principle 
cleanly extend the LRG selection of red galaxies to $\sim 0.8$ {\tt mag} lower luminosity, 
increasing the LRG comoving number density by a factor $\sim 3$ \citep{Pad04}}}.
At $z = 0.1$ (top left panel), a $\sim M^* - 0.5$ galaxy with the exact red-sequence color 
(i.e. one that lies on the solid line) would pass the LRG target selection 
(and be colored violet), but at $z = 0.2$ (bottom left), 
it would have to have an absolute magnitude $\sim M^* - 1$ to make the cut. 
The red box is the proposed rest-frame color-luminosity cut. When applied uniformly across 
cosmological epoch, we obtained a 
trimmed sample of consistent LRG population with a roughly constant comoving density (cf.
Figure \ref{fig:lrg_hist}). The slight increase in density at higher redshift is not real
but a consequence of Eddington bias from larger photometric uncertainties at high redshift.

\section{Empirical Color-Redshift Relation}\label{appendix}
Since galaxies are clustered, their neighbors tend to be at the same redshift. 
Hence, by using a set of reference galaxies (LRG or QSO) with spectroscopic redshifts, 
one can infer the average \emph{unbiased} color (and other photometric attributes) 
of a typical galaxy at those redshifts from the properties of galaxies that clustered 
around these reference galaxies, even though the selection of the original reference 
galaxies maybe biased. Specifically, for spectra obtained with the SDSS, while both 
LRGs and QSOs are color selected, and thus by design have a biased mean color, the more numerous 
galaxies from the deeper imaging data that cluster around these spectroscopic objects 
are not. By stacking fields of LRGs or QSOs with similar redshifts to make a composite 
galaxy field at those redshifts, background and foreground galaxies seen in projection 
can be removed statistically as they are not expected to correlate with the reference spectra.

We have calibrated the color-redshift relation and the \emph{observed} scatter, 
in two color dimensions ($g-r$ and $r-i$) for galaxies with a fixed 
absolute magnitude range ($M^* - 2.5 < \rpet < M^*$). An evolving model of an old 
stellar population elliptical galaxies from {\rm PEGASE} \citep{Fio97} was used to 
compute the $r$-band magnitude for a $M^*$ galaxy as a function of redshift. Two 
luminosity ranges were computed: (1) using $M^* - 2.5 < \rpet < M^*-0.8$ for $\sim M^* - 1$ 
LRG-like galaxies, (2) $M^* - 1.5 < \rpet < M^* + 0.5$ for $\sim M^*$ red sequence 
galaxies. We employ an optimal weighting using a two dimensional elliptical 
Gaussian to locate the peak and estimate the scatter of the locus. Details on 
how this is done and measurements of \emph{intrinsic} scatter of the red sequence 
to constrain the spread of formation time scale of old stars as a function of redshift, 
as well as comparison between colors of galaxies in the neighborhood of
LRGs and QSOs will be discussed in a subsequent paper \citep{Loh05}.

\begin{figure}
\includegraphics[width=0.48\textwidth]{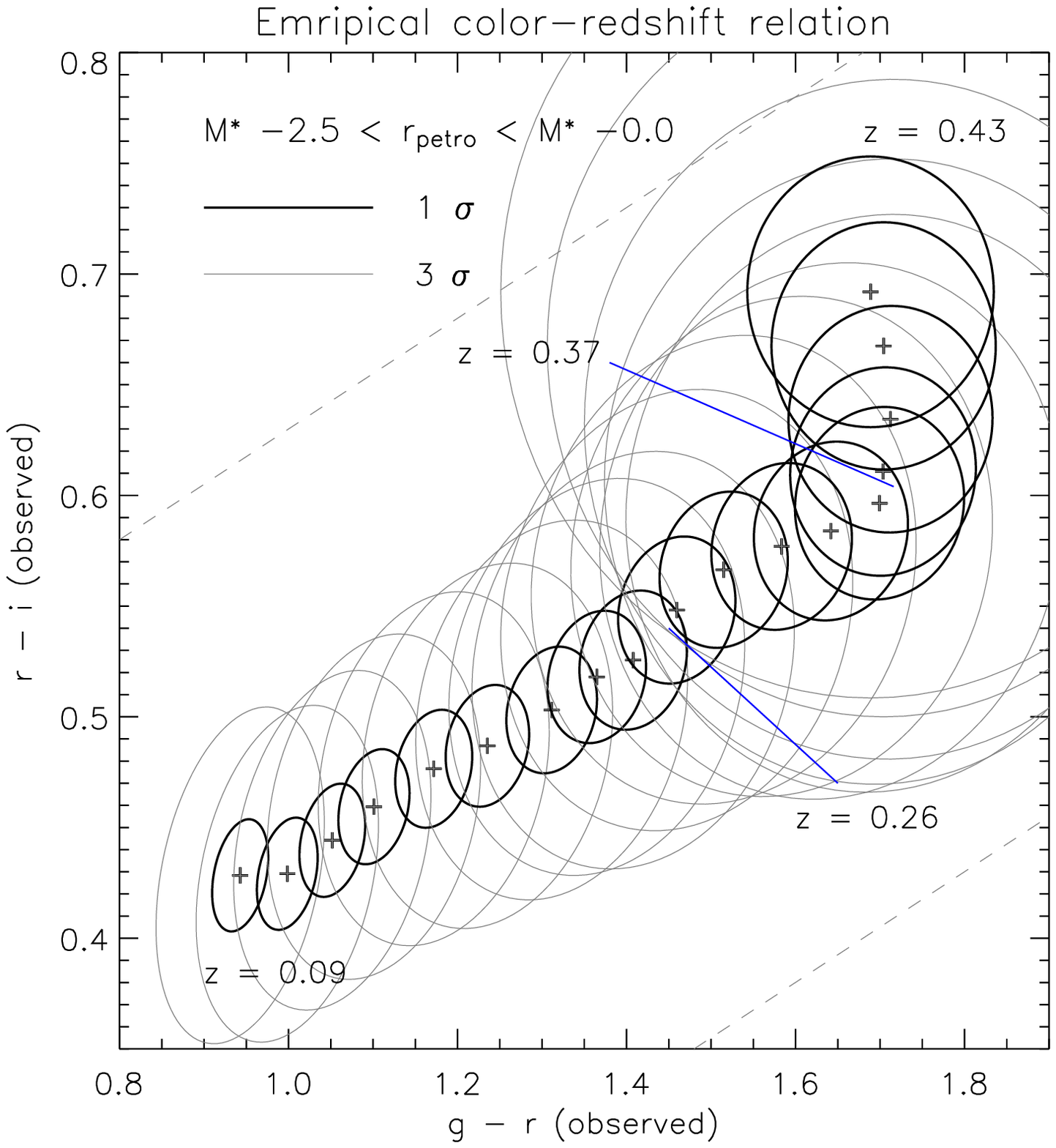}
\caption{The empirically determined locus of red sequence $M^*$ galaxies 
from $z = 0.09$ to $z = 0.43$ in increments of $\Delta z = 0.02$. The ellipses 
in black (grey) are the $1\,(3)\,\sigma$ dispersion of colors at each redshift. 
This sequence of ellipses serves as a pseudo photometric redshift filter to reduce 
contamination from background galaxies. 
\label{fig:gr_ri_locus}}
\end{figure}

Figure \ref{fig:gr_ri_locus} is the $g-r$ and $r-i$ color-color diagram of 
red-sequence galaxies with $\rpet \approx M^*$. The solid ellipses
are the $1 \sigma$ contours while the gray ellipses are $3 \sigma$.
The locus of early type galaxy follows an approximate linear sequence with redshift 
up to the up-turn point at around $z \approx 0.37$, motivating the ``Cut I'' of the
LRG target selection \citep{Eis01}. There is a slight kink at $z \approx 0.26$, 
which corresponds to the M star locus crossing with this early-type locus. The 
analysis was done using $\sim 30,000$ LRG and QSO in the redshift range 
of $0.08 < z < 0.44$ and $2340\,\sqdeg$ of DR1 imaging data processed 
with \photo 5.3 photometry.

\label{lastpage}
\end{document}